\documentclass[authoryear,3p,review]{elsarticle}
\usepackage[T1]{fontenc}
\usepackage[utf8]{inputenc}
\usepackage{bm,amsmath,amsfonts,amssymb,graphicx}
\usepackage{hyperref}
\let\today\relax
\makeatletter
\def\ps@pprintTitle{%
    \let\@oddhead\@empty
    \let\@evenhead\@empty
    \def\@oddfoot{\footnotesize\itshape
         {Submitted preprint} \hfill\today}%
    \let\@evenfoot\@oddfoot
    }
\makeatother

\begin{document}

\begin{frontmatter}
\title{Deep generative models in inversion: a review and development of a new approach based on a variational autoencoder\tnoteref{t1}}
\tnotetext[t1]{Jorge Lopez-Alvis conceived the main idea of the paper, developed and tested the proposed method (including its software implementation) and wrote a first draft of the manuscript. Eric Laloy conceived the main idea of the paper, performed some software tests, suggested improvements to the proposed method and edited the first draft. Frédéric Nguyen and Thomas Hermans are supervisors of the first author, conceived the main idea of the paper, suggested improvements to the proposed method, obtained funding for the research and edited the first draft.}

\author[address1,address2]{Jorge Lopez-Alvis\corref{mycorrespondingauthor}}
\cortext[mycorrespondingauthor]{Corresponding author}
\ead{jlopez@uliege.be}

\author[address3]{Eric Laloy}

\author[address1]{Frédéric Nguyen}

\author[address2]{Thomas Hermans}

\address[address1]{Urban and Environmental Engineering, Applied Geophysics, University of Liege, Belgium}
\address[address2]{Department of Geology, Ghent University, Belgium}
\address[address3]{Engineered and Geosystems Analysis, Institute for Environment, Health and Safety, Belgian Nuclear Research Center, Belgium}

\begin{abstract}
When solving inverse problems in geophysical imaging, deep generative models (DGMs) may be used to enforce the solution to display highly structured spatial patterns which are supported by independent information (e.g.\ the geological setting) of the subsurface. In such case, inversion may be formulated in a latent space where a low-dimensional parameterization of the patterns is defined and where Markov chain Monte Carlo or gradient-based methods may be applied. However, the generative mapping between the latent and the original (pixel) representations is usually highly nonlinear which may cause some difficulties for inversion, especially for gradient-based methods. In this contribution we review the conceptual framework of inversion with DGMs and study the principal causes of the nonlinearity of the generative mapping. As a result, we identify a conflict between two goals: the accuracy of the generated patterns and the feasibility of gradient-based inversion. In addition, we show how some of the training parameters of a variational autoencoder, which is a particular instance of a DGM, may be chosen so that a tradeoff between these two goals is achieved and acceptable inversion results are obtained with a stochastic gradient-descent scheme. A test case using truth models with channel patterns of different complexity and cross-borehole traveltime tomographic data involving both a linear and a nonlinear forward operator is used to assess the performance of the proposed approach.
\end{abstract}


\end{frontmatter}


\section{Introduction}

A common task in the geosciences is to solve an inverse problem in order to obtain a model (or image) from a set of measurements sensing a heterogeneous spatial domain. When measurements are sparse, the inverse problem is usually ill-posed and its solution non-unique. In such cases it is possible to constrain the solution to be found only among models with certain spatial patterns. In practice, such patterns are supported by independent (prior) information of the sensed domain (e.g.\ knowledge of the geological setting) and used with the aim of appropriately reconstructing heterogeneity. Classical regularization may be used to impose certain structures to the solution \citep{tikhonov_solutions_1977} but these are generally too simple to be realistic. Recently, the use of deep generative models (DGMs) to constrain the solution space of inverse problems has been proposed so that resulting models have specific spatial patterns \citep{bora_compressed_2017, laloy_inversion_2017, hand_global_2018, seo_learning-based_2019}. DGMs can deal with realistic (natural) patterns which are not captured by classical regularization or random processes defined by second-order statistics \citep{linde_geological_2015}. In this way, inversion with DGMs provides an alternative to inversion with either multiple-point geostatistics (MPS) \citep{gonzalez_seismic_2008, hansen_inverse_2012, linde_geological_2015} or example-based texture synthesis (ETS) \citep{zahner_image_2016}.

All the previously mentioned methods generally rely on gridded representations for the models (i.e.\ by dividing the spatial domain in cells or pixels). However, a key difference between them is that MPS and ETS directly extract spatial patterns from a training image (or exemplar) \citep{strebelle_conditional_2002, mariethoz_direct_2010} whereas DGMs require an initial learning phase in which samples of the patterns are used as a training set---such samples may be obtained by different means, e.g.\ cropped from a training image or taken directly from a large set of images. While any of the methods (DGMs, MPS or ETS) may be used with inversion, DGMs rely on a so called latent space where a low-dimensional parametric representation (referred to as latent vector or code) is defined and where Markov Chain Monte Carlo (MCMC) or gradient-based methods may be applied. Note that when using gridded representations each model may be seen as a vector in "pixel" space (a space where each pixel is one dimension). Since only highly structured spatial patterns are allowed, vectors will only occupy a subset of the pixel space. This subset defines a manifold of lower dimensionality than the pixel space \citep{fefferman_testing_2016} and the latent space is simply a low-dimensional space where such manifold is mapped. Most inversion methods require a perturbation step to search for models that fit the data but such a step is not straightforward to compute for highly structured patterns \citep{linde_geological_2015, hansen_inverse_2012}. The latent space of DGMs provides a useful frame to compute a perturbation step \citep{laloy_inversion_2017} or even a local gradient-descent direction \citep{laloy_gradient-based_2019} which generally results in better exploration of the posterior distribution and/or faster convergence compared to inversion with MPS or ETS.

So far, inversion with DGMs has been done successfully with regular MCMC sampling methods \citep{laloy_inversion_2017, laloy_training-image_2018}. However, when applicable, gradient-based methods may be preferred given their lower computational demand. Gradient-based deterministic inversion with DGMs has been pursued with encouraging results (Richardson, 2018, Laloy et al., 2019), however, convergence to the true model was shown to be dependent on the initial model. In the framework of probabilistic inversion, MCMC methods that use the gradient to guide the sampling in the latent space have shown to be less prone to get trapped in local minima than gradient-based deterministic methods while they are also expected to reach convergence faster than regular MCMC \citep{mosser_stochastic_2018}. A different inversion strategy that has also been applied successfully with DGMs and has a relatively low computational cost is the Ensemble Smoother \citep{canchumuni_towards_2019, mo_integration_2020}.

Recently, \citet{laloy_gradient-based_2019} studied the difficulties of performing gradient-based deterministic inversion with a specific DGM. They concluded that the non-linearity of their generative function or "generator" (i.e.\ the mapping from the latent space to the pixel space) was high enough to hinder gradient-based optimization, causing the latter to often fail in finding the global minimum even when the objective function was known to be convex (in pixel space). Such high non-linearity is expected since DGMs approximate highly complex patterns by using low-dimensional inputs usually generated by simple probability distributions (e.g.\ normal or uniform distributions). The complexity of realistic patterns means that the corresponding manifold will generally have both a curvature and a topology that is radically different from the region (or subset) defined indirectly in the latent space by the chosen probability distribution. Then, the generative function has to deform this region (when mapping it to the pixel space) in such a way as to approximate (or cover) the manifold as close as possible. The combined deformation needed to curve the region and to approximate its topology causes the generative function to be highly nonlinear. In order to approximate manifolds of realistic patterns, most common DGMs involve (artificial) neural networks with several layers and non-linear (activation) functions. Recently, it has been shown that deep neural networks with a ReLU activation function are able to change topology of the input \citep{naitzat_topology_2020} so, besides nonlinearity, the generator of DGMs may also induce changes in topology when mapping from the latent space. When the sole purpose of the DGM is for generating new samples, high nonlinearity and induced changes in topology are not important but they might be an issue when the DGM is used for additional tasks, such as inversion.

For a specific subsurface pattern, the degree of non-linearity and the changes induced in topology by the generative function may be controlled mainly by its architecture and the way it is trained \citep{goodfellow_deep_2016}. Regarding difference in training, two common types of DGMs can be distinguished: generative adversarial networks (GANs) \citep{goodfellow_generative_2014} and variational autoencoders (VAEs) \citep{kingma_auto-encoding_2014}---in both cases training generally takes the form of optimizing a loss function. GANs and VAEs require specification of a probability distribution in the latent space and an architecture for the discriminator or encoder (respectively) in addition to the one for their generators. They might also require other parameters to be specified such as the weights on the different terms of the loss function. Frequently, some of these choices use default values, but generally all of them may affect the degree of non-linearity of the generator \citep{rolinek_variational_2019}.

Given all these possible controls for learning the generator it is interesting to investigate whether there exist some combinations of such controls that allow both for a good reproduction of the patterns and a good performance of less computationally demanding gradient-based inversion. In this work, we review some of the difficulties of performing inversion with DGMs and show how to obtain a well-balanced tradeoff between accuracy in patterns and applicability of gradient-based methods. In particular, we propose to use the training choice of a VAE as DGM and to select some of its parameters in order to achieve good results with gradient-based inversion. Then, we compare this to the training choice of a GAN that has been tested with gradient-based inversion in prior studies \citep{laloy_gradient-based_2019, richardson_generative_2018}. Furthermore, we show that since the resulting VAE inversion is only mildly nonlinear, modified stochastic gradient-descent (SGD) methods are generally sufficient to avoid getting trapped in local minima and provide a better alternative than regular gradient-based methods while also retaining a low computational cost.

The remainder of this paper is structured as follows. Section \ref{Sec:DGMs} explains DGMs and their conceptualization as approximating the real (pattern) manifold. In Section \ref{Sec:InvDGMs} the use of DGMs to represent prior information in inversion and the difficulties of performing gradient-based inversion are reviewed. Sections \ref{Sec:VAEasDGM} and \ref{Sec:SGDdec} show how to use a VAE and SGD to cope with some of the mentioned difficulties. Then, Section \ref{Sec:Results} shows some results of the proposed approach. Section \ref{Sec:Discussion} discusses the obtained results and points to some remaining challenges. Finally, Section \ref{Sec:Conclusions} presents the conclusions of this work.

\section{Methods}

\subsection{Deep generative models (DGM) to represent realistic patterns.}
\label{Sec:DGMs}

The term "deep learning" generally refers to machine learning methods that involve several layers of multidimensional functions. This general "deep" setting has been shown to allow for complex mappings to be accurately approximated by building a succession of intermediate (simpler) representations or concepts \citep{goodfellow_deep_2016}. Consider, for instance, deep neural networks (DNNs) which are mappings defined by a composition of a set of (multidimensional) functions $\bm{\phi}_k$ as:

\begin{equation}
\mathbf{g}(\mathbf{x}) = (\bm{\phi}_L \circ \dots \circ \bm{\phi}_2 \circ \bm{\phi}_1)(\mathbf{x})
\end{equation} 

\noindent where $\mathbf{x}$ is a multidimensional (vector) input, $k = \{1, \dots, L\}$ denotes the function (layer) index and composition follows the order from right to left. Furthermore, each $\bm{\phi}_k$ is defined as:

\begin{equation}
\bm{\phi}_k(\bm{\xi})=\psi_k(\mathbf{M}_k\bm{\xi}+\mathbf{b}_k)
\end{equation}

\noindent in which $\psi_k$ is a (nonlinear) activation function, $\mathbf{M}_k$ and $\mathbf{b}_k$ are vectors of weights and biases, respectively, and $\bm{\xi}$ denotes the output of the previous function (layer) $\bm{\phi}_{k-1}$ for $k>1$ or the initial input $\mathbf{x}$ for $k = 1$. Then, training the DNN involves estimating the values for the all the parameters $\bm{\theta}=\{\mathbf{M}_k,\mathbf{b}_k \mid 1 \leq k \leq L\}$ where each $\mathbf{M}_k$ or $\mathbf{b}_k$ may be of different dimensionality depending on the layer. In practice, the number of parameters $\bm{\theta}$ for such models may reach the order of $10^6$, therefore training is achieved by relying on autodifferentiation \citep[see e.g.][]{paszke_automatic_2017} and fast optimization techniques based on stochastic gradient descent (SGD) \citep[see e.g.][]{kingma_adam_2017}, both usually implemented for and run in highly parallel (GPU) computing architectures.

\begin{figure*}
\includegraphics[width=\textwidth]{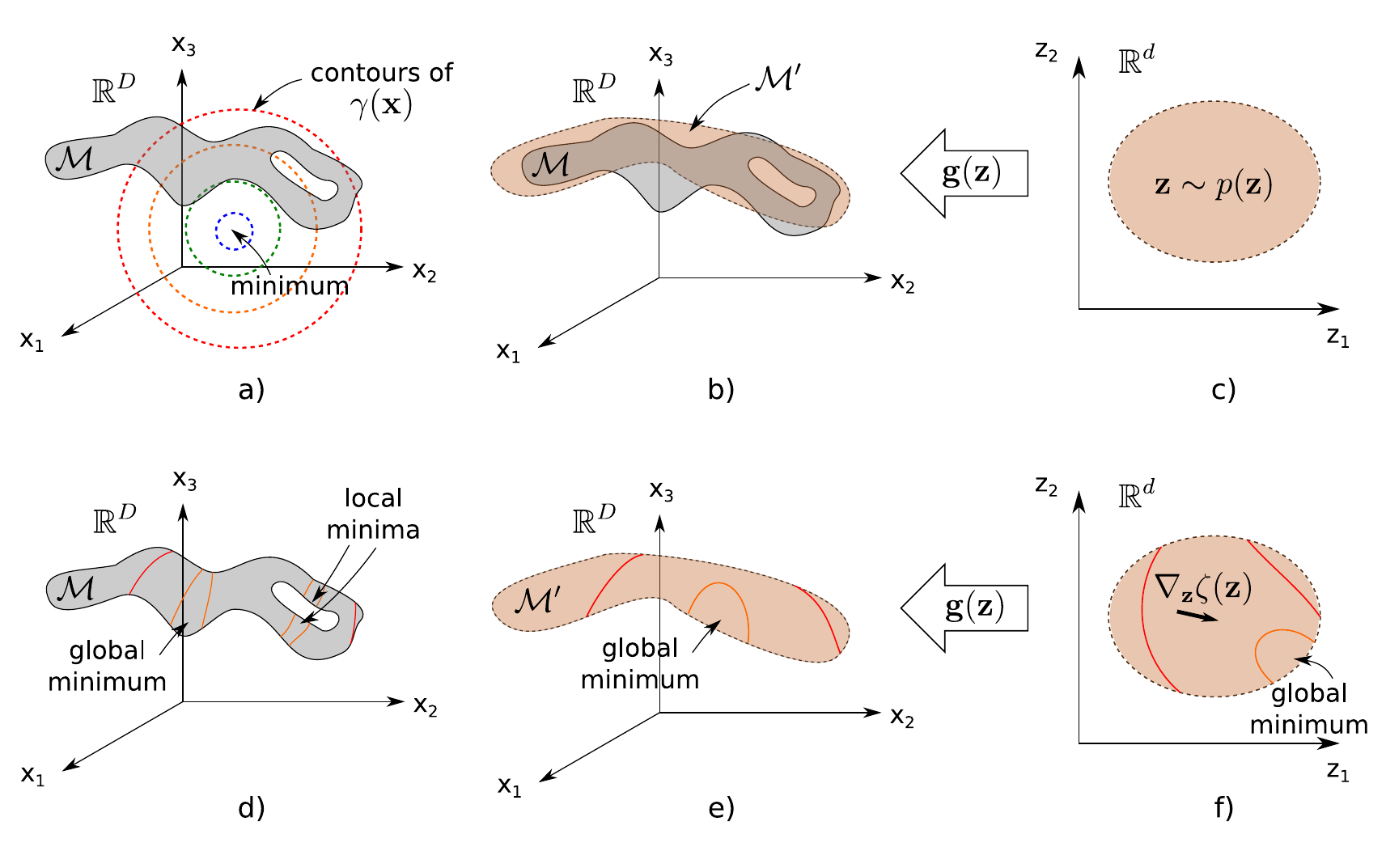}
\centering
\caption{Sketch of low-dimensional manifold setting. (a) Real manifold $\mathcal{M}$ and misfit function $\gamma(\mathbf{x})$ in ambient space $\mathbb{R}^D$. (b) Approximate manifold $\mathcal{M}'$ overlaying the real manifold. (c) Region of latent space $\mathbb{R}^d$ where the approximate manifold is implicitly defined by the probability distribution $p(\mathbf{z})$. (d) Misfit function contours intersected by the real manifold. (e) Misfit function contours intersected by approximate manifold. (f) Misfit function contours back-mapped onto the latent space and the related gradient $\nabla_\mathbf{z}\zeta(\mathbf{z})$ computed at one iteration.}
\label{Fig:manifold}
\end{figure*}

A deep generative model (DGM) is a particular application of such deep methods \citep{salakhutdinov_learning_2015}. In a DGM a set of training examples $\mathbf{X}=\{\mathbf{x}^{(i)} \mid 1 \leq i \leq N\}$ and a simple low-dimensional probability distribution $p(\mathbf{z})$ are used to learn a model $\mathbf{g}(\mathbf{z})$ that is capable of generating new samples of $\mathbf{x}$ (which are consistent with the training set) by using as input samples from $p(\mathbf{z})$. This can be written as:

\begin{equation}
\mathbf{x}=\mathbf{g}(\mathbf{z}),\; \mathbf{z} \sim p(\mathbf{z})
\end{equation}

\noindent where $\mathbf{g}(\mathbf{z})$ is referred to as the "generator" and $\mathbf{z}$ denotes a vector of latent variables or "code". While the training (and generated) samples $\mathbf{x}$ are usually represented in a high-dimensional space $\mathbb{R}^D$, the probability distribution $p(\mathbf{z})$ is defined in a low-dimensional space $\mathbb{R}^d$. The space $\mathbb{R}^D$ is often referred to as "ambient space" while the space $\mathbb{R}^d$ is called the "latent space". Fig.\ \ref{Fig:manifold}a,c shows a schematic representation where the dimensionality of the ambient space is $D=3$  and the one of the latent space is $d=2$. A typical application of DGMs is the generation of images \citep[see e.g.][]{kingma_auto-encoding_2014, goodfellow_generative_2014} for which the ambient space is just the pixel space. Gridded representations of subsurface models may be seen as two- or three-dimensional images of the subsurface.

The underlying assumption in DGMs is that real-world data are generally structured in their high-dimensional ambient space $\mathbb{R}^D$ and therefore have an intrinsic lower dimensionality---such assumption is known in machine learning literature as the manifold hypothesis \citep{fefferman_testing_2016} because it states that high-dimensional data usually lie on (or lie close to) a lower-dimensional manifold $\mathcal{M} \subset \mathbb{R}^D$. For instance, when studying a subsurface region it is usually assumed that geological processes gave it certain degree of structure then, to allow for a flexible base on which to represent the distribution of the different subsurface materials, the region is usually divided in pixels (or cells) within each of which the material is assumed to be homogeneous. Such gridded representation "lives" in the high-dimensional pixel space (the ambient space) but since it has some structure there should be a lower dimensional space (the latent space) where the same distribution of subsurface materials might be represented. Technically, while both the latent space $\mathbb{R}^d$ and the manifold $\mathcal{M}$ are usually low-dimensional, they may differ in dimensionality and/or the manifold may only occupy a certain portion of the latent space (e.g.\ the shaded region in Fig.\ \ref{Fig:manifold}c).

Considering the manifold assumption described above, a DGM may be regarded as a model to implicitly approximate the "real" manifold $\mathcal{M}$ by generating samples that closely follow such manifold, i.e.\ that lie on an approximate manifold $\mathcal{M}'$ (Fig.\ \ref{Fig:manifold}b). Samples of this approximate manifold are generated by sampling first from a simple probability distribution $p(\mathbf{z})$ in latent space (e.g.\ a normal or uniform distribution) and then passing them through the generator $\mathbf{g}(\mathbf{z})$. Since the probability distribution $p(\mathbf{z})$ defines indirectly a region (or subset) in latent space that generally has a different curvature and topology than the real manifold, the generator $\mathbf{g}(\mathbf{z})$ must be able to approximate both curvature and topology when mapping the samples of $p(\mathbf{z})$ to ambient space. This generally requires the generator to be a highly nonlinear function. As an instance, consider the case of certain spatial patterns whose real manifold is a highly curved surface with "holes" in ambient space and the (input) region defined by a uniform $p(\mathbf{z})$ is a (flat) plane in a two-dimensional latent space. Regarding their topological properties, one technically says that this plane is simply connected while the real manifold is not \citep[see e.g.][]{kim_data_2019}. Then, the generative function has to deform this plane in such a way as to approximate (or cover) the real manifold as close as possible. An important property of DGMs is that since a probability distribution in latent space is used, the sample "density" of such plane (and its mapping) also plays an important role. For instance, the generative function may approximate the "holes" of the real manifold by creating regions of very low density of samples when mapping to ambient space (to picture this one can imagine locally stretching a flexible material without changing its curvature). The combined deformation needed to curve the plane and to "make" the holes causes the generative function to be highly nonlinear. Note that when considering a DGM that uses a DNN with ReLU activation functions as generator $\mathbf{g}(\mathbf{z})$, it is also possible for $\mathbf{g}(\mathbf{z})$ to change topology of the input by "folding" transformations \citep{naitzat_topology_2020}.

While one should always strive to accurately approximate the real manifold, since a finite set of training samples is used a tradeoff between accuracy and diversity in the generated samples may be a better objective. Indeed, the use of the prescribed probability distributions is done to continuously "fill" the space between the samples and therefore generate samples of a continuous manifold. Recent success---in terms of accuracy and diversity of generated samples---has been achieved with two DGMs that are based on deep neural networks (DNNs): generative adversarial networks (GANs) \citep{goodfellow_generative_2014} and variational autoencoders (VAEs) \citep{kingma_auto-encoding_2014}. The generator $g(\mathbf{z})$ on both strategies is a mapping from low-dimensional input $\mathbf{z} \in \mathbb{R}^d$ to high-dimensional output $\mathbf{x} \in \mathbb{R}^D$. In contrast, the mappings corresponding to the discriminator and the encoder take high-dimensional inputs $\mathbf{x}$ and return low-dimensional outputs.

\subsection{Gradient-based inversion with DGMs}
\label{Sec:InvDGMs}

Consider a survey or experiment for which a vector of noisy measurements $\mathbf{y} = (y_1,\dots,y_Q)^T \in \mathbb{R}^Q$ of a physical process is available. A simplified description of the process may be expressed by a (mathematical) forward operator $\mathbf{f}:\mathbb{R}^D \to \mathbb{R}^Q$ that takes as input a subsurface model vector $\mathbf{x} = (x_1,\dots,x_D)^T \in \mathbb{R}^D$ (obtained by discretizing the spatial distribution of physical properties) and outputs a simulated response $\mathbf{f}(\mathbf{x})$. Commonly, this operator is in the form of a (numerical) discretization of a set of PDEs describing the process under study and is an approximation of the real process. As a result of such approximation and the use of noisy data, a noise term $\bm{\eta}$ is added to the simulation to represent total uncertainty. Then, the relation between the operator and the measurements may be written as \citep[see e.g.][]{aster_parameters_2013}:

\begin{equation}
\mathbf{y}=\mathbf{f}(\mathbf{x})+\bm{\eta}
\label{Eq:inversion}
\end{equation}

\noindent The corresponding inverse problem or inversion of Eq.\ (\ref{Eq:inversion}), aims to obtain an estimation of the vector $\mathbf{x}$ from the (noisy) data $\mathbf{y}$. Deterministic inversion does so by optimizing a misfit function $\gamma(\mathbf{x})$ that is usually given in the form of a distance function between simulated response $\mathbf{f}(\mathbf{x})$ and data $\mathbf{y}$, e.g.\ by the $l_2$ norm:

\begin{equation}
\gamma(\mathbf{x}) = \|\mathbf{f}(\mathbf{x})-\mathbf{y}\|^2
\label{Eq:objx}
\end{equation}

\noindent In this way, traditional gradient-based inversion requires the gradient $\nabla_\mathbf{x}\gamma(\mathbf{x})$ whose elements are:

\begin{equation}
[\nabla_\mathbf{x}\gamma(\mathbf{x})]_i = \frac{\partial \gamma(\mathbf{x})}{\partial x_i}
\label{Eq:gradobj}
\end{equation}

\noindent and are computed by considering Eq.\ (\ref{Eq:inversion}) together with the chosen misfit.

DGMs may be used with inversion of subsurface data $\mathbf{y}$ to obtain geologically realistic spatial distributions of physical properties $\mathbf{x}$ \citep{laloy_inversion_2017}. While this is also possible with traditional deterministic inversion where a regularization term is added directly in Eq.\ (\ref{Eq:objx}) (i.e.\ in ambient space) to obtain models with the imposed structures that minimize the misfit \citep{lange_frequency_2012, caterina_case_2014}, DGMs are more flexible because they can enforce different structures provided they are appropriately trained. In the DGM setting, the low-dimensional samples $\mathbf{z}$ that input to the generator $\mathbf{g}(\mathbf{z})$ may be seen as defining a low-dimensional parameterization (or encoding) of realistic patterns $\mathbf{x}$ and therefore exploration of the set of feasible models may be done in the latent space $\mathbb{R}^d$, as long as the search is done within the region where the approximated manifold $\mathcal{M}'$ is defined (depicted by shading in Fig.\ \ref{Fig:manifold}c).

Since the misfit $\gamma(\mathbf{x})$ is typically defined in ambient space $\mathbb{R}^D$ (e.g.\  in Fig.\ \ref{Fig:manifold}a), gradient-based inversion with DGMs may be seen as optimizing the intersection of $\gamma(\mathbf{x})$ with the approximate manifold $\mathcal{M}'$ (Fig.\ \ref{Fig:manifold}e). Such intersected misfit is mapped into the latent space (Fig.\ \ref{Fig:manifold}f) and may be expressed as $\gamma(\mathbf{g}(\mathbf{z}))$. Also note that when probability distributions $p(\mathbf{z})$ with infinite support are used (e.g.\ a normal distribution), one can guide the search in the latent space by adding controlling (regularization) terms to the mapped misfit \citep[see e.g.][]{bora_compressed_2017} and the resulting objective function may be written as:

\begin{align}
\zeta(\mathbf{z}) &= \gamma(\mathbf{g}(\mathbf{z})) + \lambda R(\mathbf{z}) \nonumber \\
&= \|\mathbf{f}(\mathbf{g}(\mathbf{z}))-\mathbf{y}\|^2 + \lambda R(\mathbf{z})
\label{Eq:objz}
\end{align}

\noindent where $R(z)$ is a regularization term defined in the latent space and $\lambda$ is the corresponding regularization factor. 

In practice, no exhaustive mapping has to be done and the gradient $\nabla_\mathbf{z}\zeta(\mathbf{z})$ is only computed for the points in latent space where optimization lands in each iteration (in Fig.\ \ref{Fig:manifold}f the gradient is represented for one iteration). The gradient $\nabla_\mathbf{z}\zeta(\mathbf{z})$ is computed by adding a derivative layer corresponding to $\nabla_\mathbf{x}\gamma(\mathbf{x})$ to the autodifferentation that was set up for $\mathbf{g}(\mathbf{z})$ while training the DGM \citep[see e.g.][]{laloy_gradient-based_2019}. Such autodifferentiation setup may be seen as implicitly obtaining the jacobian $\mathbf{J}(\mathbf{z})$ of size $D \times d$ whose elements are:

\begin{equation}
[\mathbf{J}({\mathbf{z}})]_{i,j} = \frac{\partial g_i(\mathbf{z})}{\partial z_j}
\label{Eq:jacobian}
\end{equation}

\noindent Then, the gradient $\nabla_\mathbf{z}\zeta(\mathbf{z})$ is obtained from Eq.\ (\ref{Eq:objz}) by using the chain rule given by the product of Eqs.\ (\ref{Eq:gradobj}) and (\ref{Eq:jacobian}):

\begin{align}
\nabla_\mathbf{z}\zeta(\mathbf{z}) &= \nabla_\mathbf{z}\gamma(\mathbf{g}(\mathbf{z})) + \lambda \nabla_\mathbf{z} R(\mathbf{z}) \nonumber \\
&= \mathbf{J}(\mathbf{z})^T \nabla_\mathbf{x}\gamma(\mathbf{x}) + \lambda \nabla_\mathbf{z} R(\mathbf{z})
\label{Eq:gradz}
\end{align}

\noindent The latter may also be done implicitly by incorporating directly in the autodifferentiation framework, e.g. putting it on top of the so called computational graph \citep{richardson_generative_2018, mosser_stochastic_2018}.

Assuming the considered misfit function $\gamma(\mathbf{x})$ is convex in ambient space $\mathbb{R}^D$ (as depicted by concentric contours in Fig.\ \ref{Fig:manifold}a), difficulties to perform gradient-based deterministic inversion may arise due to the generator $\mathbf{g}(\mathbf{z})$ \citep{laloy_gradient-based_2019}. We propose that such difficulties arise because the generator (1) is highly nonlinear and (2) changes the topology of the input region defined by $p(\mathbf{z})$. Both of these properties often cause distances (between samples) in latent space to be significantly different than distances in ambient space. Consider again the example of a real manifold that is a highly curved surface with "holes" in it and a uniform distribution $p(\mathbf{z})$ is used as input to the generator, then the latter might be able to approximate both the curvature and the holes at the cost of increasing nonlinearity and/or changing topology. When considering this backwards---e.g.\ when mapping the misfit function $\gamma(\mathbf{x})$ in the latent space---the approximation of both high curvature and differences in topology often translate in discontinuities or high nonlinearities because a continuous mapping onto the uniform distribution is enforced. This results in high curvature being effectively "flattened" and holes effectively "glued", both of which cause distances to be highly distorted. In this work, we will call a generator "well-behaved" when it is only mildly nonlinear and preserves topology.

Both the generator's nonlinearity and its ability to change topology, may be controlled by two factors: (1) the generator architecture (type and size of each layer and total number of layers) and (2) the way it is trained (including training parameters). If the goal is to perform gradient-based inversion with DGMs, one should try to preserve convexity of $\gamma(\mathbf{x})$ as much as possible when mapping it to the latent space as $\gamma(\mathbf{g}(\mathbf{z}))$ while not degrading the generator's ability to reproduce the desired patterns. To aid in preserving such convexity, we propose to enforce the generator $\mathbf{g}(\mathbf{z})$ to be well-behaved. This means that the generator will approximate the real manifold $\mathcal{M}$ with a manifold $\mathcal{M}'$ with a moderate curvature and whose topology is the same as the region defined in latent space by $p(\mathbf{z})$. By enforcing a moderate curvature manifold, local oscillations that may give rise to local minima (as those shown in Fig.\ \ref{Fig:manifold}d) but only have minimum impact in pattern accuracy are avoided in the approximate manifold $\mathcal{M}'$ (the local minima are no longer present in Fig.\ \ref{Fig:manifold}e). In turn, when the generator is encouraged to preserve topology no more local minima should arise in $\mathbb{R}^d$ than the ones resulting from intersecting $\gamma(\mathbf{x})$ with the approximate manifold $\mathcal{M}'$ in $\mathbb{R}^D$ (note e.g.\ there is one local minima in both Fig.\ \ref{Fig:manifold}e,f). The latter is in line with the proposal of \citet{falorsi_explorations_2018}, where they argue that for the purpose of representation learning (which basically means learning encodings that are useful for other tasks than just generative modeling) the mapping should preserve topology.

GANs often produce highly nonlinear generators that do not preserve topology, which may result in challenging inversion in the latent space. \citet{laloy_training-image_2018} provide an example of how architecture of a GAN is set to obtain a relatively well-behaved generator $\mathbf{g}(\mathbf{z})$. They propose to use a model called spatial generative adversarial network (SGAN) \citep{jetchev_texture_2017} that enforces different latent variables to affect different local regions in the ambient space. Their architecture results in a high compression (lower dimensionality of the latent space) and controls nonlinearity which allowed them to successfully perform MCMC-based inversion in the latent space. However, gradient-based deterministic inversion performed with the same DGM was shown to be highly dependent on the initial model \citep{laloy_gradient-based_2019} pointing towards the existence of local minima. In this work we aim for robust gradient-based inversion in latent space by considering a VAE, the other predominant type of DGM, and its ability to produce a well-behaved generator.

\subsection{Variational autoencoder (VAE) as DGM for inversion}
\label{Sec:VAEasDGM}

A variational autoencoder (VAE) is the model resulting from using a reparameterized gradient estimator for the evidence lower bound while applying (amortized) variational inference to an autoencoder, i.e.\ an architecture involving an encoder and a decoder which are both (possibly deep) neural networks \citep{kingma_auto-encoding_2014, zhang_advances_2018}. To train a VAE one uses a dataset $\mathbf{X}=\{\mathbf{x}^{(i)} \mid 1 \leq i \leq N\}$ where each $\mathbf{x}^{(i)}$ is a sample (e.g.\ an image) with the desired patterns and then maximizes the sum of the evidence (or marginal likelihood) lower bound of each individual sample. The evidence lower bound for each sample can be written as \citep{kingma_auto-encoding_2014}

\begin{equation}
\mathcal{L}(\theta,\vartheta;\mathbf{x}^{(i)})=\mathcal{L}^x + \mathcal{L}^z
\label{Eq:ELBO}
\end{equation}

\noindent with

\begin{equation}
\mathcal{L}^x = \mathbb{E}_{q_\vartheta(\mathbf{z}|\mathbf{x}^{(i)})}[\log(p_\theta(\mathbf{x}^{(i)}|\mathbf{z})]
\label{Eq:Lx}
\end{equation}

\noindent and

\begin{equation}
\mathcal{L}^z=-D_{KL}(q_\vartheta(\mathbf{z}|\mathbf{x}^{(i)})||p(\mathbf{z}))
\label{Eq:Lz}
\end{equation}

\noindent where $\mathbf{z}$ refers to the codes or latent vectors, $p_\theta(\mathbf{x}|\mathbf{z})$ is the (probabilistic) decoder, $q_\vartheta(\mathbf{z}|\mathbf{x})$ is the (probabilistic) encoder, $\mathbb{E}$ denotes the expectation operator, $D_{KL}$ denotes the Kullback-Leibler distance and, $\theta$ and $\vartheta$ are the parameters (weights and biases) of the DNNs for the decoder and encoder, respectively.

In order to maximize the evidence lower bound in Eq. (\ref{Eq:ELBO}), its gradient with respect to both $\theta$ and $\vartheta$ is required, however, this is generally intractable and therefore an estimator is used.  This estimator is based on a so called reparameterization trick of the random variable $\widetilde{\mathbf{z}}\sim q_\vartheta(\mathbf{z}|\mathbf{x})$ which uses an auxiliary noise $\bm{\epsilon}$. In the case of a VAE, the encoder is defined as a multivariate Gaussian with diagonal covariance:

\begin{equation}
q_\vartheta(\mathbf{z}|\mathbf{x}) = \mathcal{N}(\mathbf{h}_\vartheta(\mathbf{x}),\mathbf{u}_\vartheta(\mathbf{x}) \cdot I_d)
\end{equation}

\noindent where $\mathbf{h}_\vartheta(\mathbf{x})$ and $\log\mathbf{u}_\vartheta(\mathbf{x})$ are modeled with DNNs and $I_d$ is a $d \times d$ diagonal matrix. Then, the encoder and the auxiliary noise $\bm{\epsilon}$ are used in the following way during training \citep{kingma_auto-encoding_2014}

\begin{equation}
\widetilde{\mathbf{z}} = \mathbf{h}_\vartheta(\mathbf{x})+\mathbf{u}_\vartheta(\mathbf{x}) \odot \bm{\epsilon},\;  \bm{\epsilon} \sim p(\bm{\epsilon})
\label{Eq:pepsilon}
\end{equation}

\noindent where $\odot$ denotes an element-wise product. Often Eq.\ (\ref{Eq:Lz}) has an analytical solution, then only Eq.\ (\ref{Eq:Lx}) is approximated with the estimator as \citep{kingma_auto-encoding_2014}

\begin{equation}
\widetilde{\mathcal{L}}^x = \frac{1}{M}\sum_{j=1}^M \log(p_\theta(\mathbf{x}^{(i)}|\widetilde{\mathbf{z}}^{(i,j)}))
\label{Eq:Lxest}
\end{equation}

\noindent where $\widetilde{\mathbf{z}}^{(i,j)} = \mathbf{h}_\vartheta(\mathbf{x}^{(i)})+\mathbf{u}_\vartheta(\mathbf{x}^{(i)}) \odot \bm{\epsilon}^{(j)}$, $\bm{\epsilon}^{(j)} \sim p(\bm{\epsilon})$ and $M$ is the number of samples used for the estimator. Further, if we set the decoder $p_\theta(\mathbf{x}|\mathbf{z})$ as a multivariate Gaussian with diagonal covariance structure, then

\begin{equation}
p_\theta(\mathbf{x}|\mathbf{z}) = \mathcal{N}(\mathbf{g}_\theta(\mathbf{z}),\mathbf{v}_\theta(\mathbf{z}) \cdot I_D)
\end{equation}

\noindent where $\mathbf{g}_\theta(\mathbf{z})$ and $\log\mathbf{v}_\theta(\mathbf{z})$ are modeled with DNNs and $I_D$ is a $D \times D$ diagonal matrix. In this work, we consider only the mean of the decoder $p_\theta(\mathbf{x}|\mathbf{z})$ which is just the (deterministic) generator $\mathbf{g}_\theta(\mathbf{z})$. Then, the corresponding (mean-square error) loss function may be written as

\begin{equation}
\widetilde{\mathcal{L}}^x = \frac{1}{M}\sum_{j=1}^M \|\mathbf{g}_\theta(\widetilde{\mathbf{z}}^{(i,j)})-\mathbf{x}^{(i)}\|^2
\label{Eq:Lxloss}
\end{equation}

\noindent The described setting allows for the gradient to be computed with respect to both $\theta$ and $\vartheta$ and then stochastic gradient descent is used to maximize the lower bound in Eq.\ (\ref{Eq:ELBO}). In the rest of this work, we drop the subindex $\theta$ in $\mathbf{g}(\mathbf{z})$ to simplify notation and also because once the DGM is trained, the parameters $\theta$ do not change, i.e.\ they are fixed for the subsequent inversion.

As previously mentioned, it is often possible to analytically integrate the Kullback-Leibler distance in Eq.\ (\ref{Eq:Lz}). In this work, we consider that $p(\mathbf{z})$ and $q_\vartheta(\mathbf{z}|\mathbf{x})$ are both Gaussian therefore Eq.\ (\ref{Eq:Lz}) may be rewritten as \citep{kingma_auto-encoding_2014}:

\begin{equation}
\mathcal{L}^z=\frac{1}{2} \sum_{i=1}^d (1+\log((u_i)^2)-(h_i)^2-(u_i)^2)
\label{Eq:Lzloss}
\end{equation}

\noindent where the sum is done for the $d$ output dimensions of the encoder.

\begin{figure}
\includegraphics{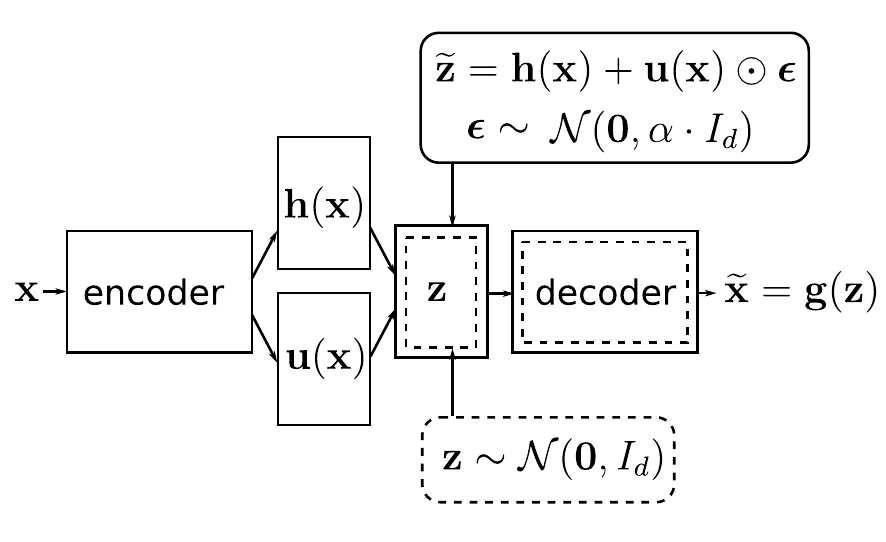}
\centering
\caption{A diagram for a VAE: steps needed for training are shown in frames with continuous line and steps needed for generation are in frames with dashed lines.}
\label{Fig:vae_diagram}
\end{figure}

Note that the term in Eqs.\ (\ref{Eq:Lx}), (\ref{Eq:Lxest}) and (\ref{Eq:Lxloss}) may be interpreted as a reconstruction term that causes the outputs of the encode-decode operation to look similar to the training samples, while the term in Eqs.\ (\ref{Eq:Lz}) and (\ref{Eq:Lzloss}) may be considered a regularization term that enforces the encoder $q_\vartheta(\mathbf{z}|\mathbf{x})$ to be close to a prescribed distribution $p(\mathbf{z})$. In practice, one may add a weight to the second term \citep{higgins_beta-vae:_2017} of the lower bound as:

\begin{equation}
\widetilde{\mathcal{L}}(\theta,\vartheta;\mathbf{x}^{(i)})=\widetilde{\mathcal{L}}^x + \beta \mathcal{L}^z
\label{Eq:aproxELBO}
\end{equation}

\noindent to prevent samples to be encoded far from each other in the latent space, which may cause overfitting of the reconstruction term and degrade the VAE's generative performance. The overall process of training and generation for a VAE is depicted in Fig. \ref{Fig:vae_diagram}.

Note that in setting up the VAE one has to choose: (1) the architectures of the encoder and decoder, (2) the probability distribution $p(\mathbf{z})$, (3) the noise distribution $p(\bm{\epsilon})$ and (4) the regularization weight $\beta$. As mentioned in Section \ref{Sec:InvDGMs}, these choices may impact the nonlinearity of the generator and its ability to preserve topology, which in turn affect the mapping of the data misfit function $\gamma(\mathbf{x})$ in latent space and possibly diminish the performance of inversion methods. In this work, we assume that an architecture for the generator (decoder) $\mathbf{g}(\mathbf{z})$ is chosen so that it performs sufficiently good in terms of reproducing the patterns. For instance, when gridded spatial distributions (images) are considered, a typical choice is a deep-convolutional neural network \citep{radford_unsupervised_2016}. While the choice of the probability distribution $p(\mathbf{z})$ may aid in obtaining a well-behaved generator, e.g.\ by selecting a probability distribution with the same (or similar) topology as the real manifold \citep{falorsi_explorations_2018}, we expect such a choice to be highly problem (pattern) dependent. Therefore herein we focus on the other two possible controls: the noise distribution $p(\bm{\epsilon})$ and the regularization weight $\beta$. 

The effect of the regularization weight $\beta$ is such that when increased the encoded training samples tend to lie closer to the prescribed probability distribution $p(\mathbf{z})$. Then, one may picture the transformation of the encoder as taking the low-dimensional approximate manifold in the ambient space and charting it (e.g. by bending, stretching and even folding) into the region defined by $p(\mathbf{z})$ in the latent space and the generator as the transformation undoing such charting.

While the effect of $\beta$ in a VAE is relatively easy to understand, the effect of the noise distribution $p(\bm{\epsilon})$ is not so straightforward. First, note that the typical choice of a diagonal noise as $p(\bm{\epsilon}) = \mathcal{N}(\mathbf{0}, \alpha \cdot I_d)$ where $\alpha$ denotes a constant variance (frequently set to $\alpha=1.0$) and $I_d$ is a $d \times d$ diagonal matrix is usually done for tractability or computational convenience \citep{kingma_auto-encoding_2014, rolinek_variational_2019}. However, it has been proposed recently that the choice of a diagonal noise has an impact on a property called disentanglement \citep{rolinek_variational_2019}. Such disentanglement basically means that different latent directions control different independent characteristics of the training (or generated) samples. They explain that a diagonal $p(\bm{\epsilon})$ might induce an encoding that preserves local orthogonality of the ambient space. In this work, we argue that the choice of a diagonal $p(\bm{\epsilon})$ (which is usually done only for computational convenience) might be useful in producing a well-behaved generator. 

\begin{figure}
\includegraphics{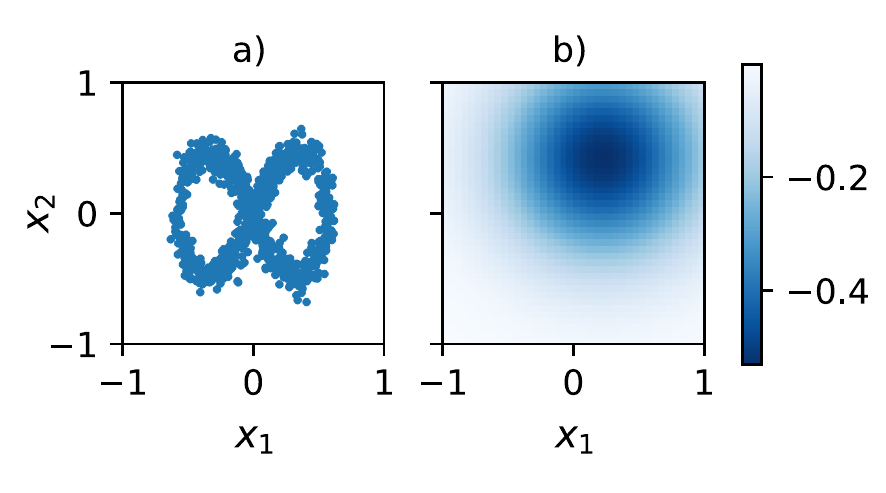}
\centering
\caption{Synthetic example of two-dimensional "eight-shaped" manifold: (a) training samples lying close the manifold, and (b) synthetic misfit function $\gamma(\mathbf{x})$.}
\label{Fig:trueobj}
\end{figure}

\begin{figure*}
\includegraphics[width=0.9\textwidth]{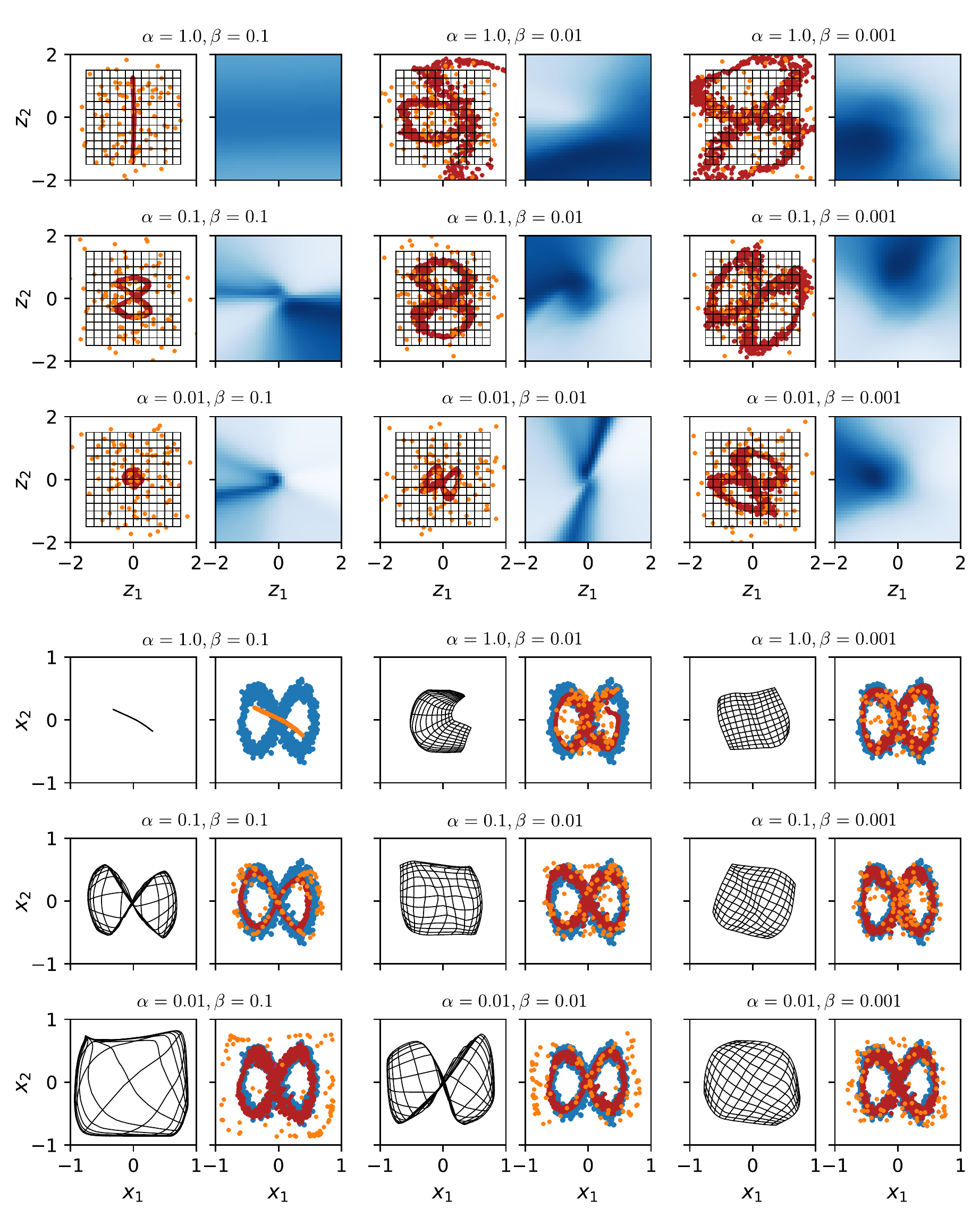}
\centering
\caption{Mapping a region of the latent space by the generator $\mathbf{g}(\mathbf{z})$ and mapping of the misfit function $\gamma(\mathbf{x})$ to the latent space with different values for $\alpha$ and $\beta$. The first three rows (z-axes) depict the latent space where each case shows: (left frame) orthogonal grid (black), encoded training samples (red) and generated samples (orange); (right frame) misfit function mapped in latent space (blue). The last three rows (x-axes) depict the ambient space where each case shows: (left frame) the same grid but mapped by the generator; (right frame) training (blue), reconstructed (red) and generated samples (orange).}
\label{Fig:deformed}
\end{figure*}

In order to visualize the joint effect of $\alpha$ and $\beta$, Fig.\ \ref{Fig:trueobj} shows a synthetic example where samples in a two-dimensional ambient space lie close to a rotated "eight-shaped" manifold (Fig.\ \ref{Fig:trueobj}a). In addition, to study the impact on inversion, a convex data misfit function $\gamma(\mathbf{x})$ in the same space (created synthetically with a negative isotropic Gaussian function) is shown in Fig.\ \ref{Fig:trueobj}b. The latent space is also chosen two-dimensional for visualization purposes but recall that for a real case the dimensionality of the latent space is usually much lower than the one of the ambient space. Then, Fig.\ \ref{Fig:deformed} considers nine different combinations for the values of $\alpha$ and $\beta$ to show how the (nonlinear) generator $\mathbf{g}(\mathbf{z})$ maps a region of the latent space (denoted by the z-axes in the first three rows) into the ambient space (denoted by the x-axes in the last three rows) in order to approximate the manifold in Fig.\ \ref{Fig:trueobj}a. To visualize the deformation caused by the generator, an orthogonal grid in the z-axes and its mapping into the x-axes (a deformed grid) are shown (both on the left of each inset). The corresponding encoded training samples are shown in red in the z-axes (left of each inset) and their reconstruction (resulting from the operation of encode-decode) is shown also in red in the the x-axes (right of each inset), where also the original training samples are shown (in blue) to assess the accuracy of reconstruction. Samples obtained from a Gaussian distribution with a unitary diagonal covariance $p(\mathbf{z})$ are shown in the z-axes in orange (left of each inset), while their generator-mapped values are shown also in orange in the x-axes (right of each inset). Finally, the mapping of the data misfit function in Fig.\ \ref{Fig:trueobj}b into the latent space is shown in the z-axes (right of each inset).

It is worth mentioning a few effects visible in the illustrative example of Figs.\ \ref{Fig:trueobj} and \ref{Fig:deformed}. First, note that increasing the variance of $\alpha$ seems to cause the grid to be more "rigid" locally (grid lines tend to intersect more at right angles) while going through the generator which may in turn help in preserving topology and controlling non-linearity (e.g.\ compare the deformation of the grids for different values of $\alpha$ for $\beta=0.01$), and more importantly, in preserving the convexity of the data misfit function in the latent space (the mapped misfit function using $\alpha=0.1$ and $\beta=0.01$ has a single global minimum, while the misfit function for $\alpha=0.01$ and $\beta=0.01$ has two minima in latent space). Also note that both $\alpha$ and $\beta$ should be set in order to not cause a significant degradation in: (1) the reconstruction of the patterns, e.g.\ the cases of $\alpha=1.0$ with both $\beta=0.1$ and $\beta=0.01$ show that the "eight-shape" is not completely reconstructed (seen in red samples not fully overlaying the blue samples in x-axes), or (2) the similarity of the encoded samples to the prescribed distribution $p(\mathbf{z})$, e.g.\ the case of $\alpha=0.01$ and $\beta=0.1$ shows that encoded samples (in red) are too concentrated (lower variance) and therefore far from the prescribed normal distribution with unit variance. In this case, the intermediate values ($\alpha=0.1$ and $\beta=0.01$) seem to provide the best choice in terms of reconstruction of the patterns, generative accuracy and convexity of the misfit function in latent space.

In summary, a generator $\mathbf{g}(\mathbf{z})$ that preserves topology and contains nonlinearity is the best choice for gradient-based inversion in the latent space because it preserves convexity of the objective function. Note, however, that if the topology of the probability distribution $p(\mathbf{z})$ is different to the one of the real manifold $\mathcal{M}$, this strategy may result in approximate manifolds $\mathcal{M}'$ that do not account for all topological differences---e.g.\ that partially cover holes of the real one (see e.g.\ Fig.\ \ref{Fig:manifold}b)---and therefore might produce models that have non-accurate patterns when sampling from $p(\mathbf{z})$. We argue that the two training parameters $\alpha$ and $\beta$ of a VAE may be chosen in order for the latter issue to not be severe, i.e.\ the generated patterns do not deviate too much from the training patterns, while still approximately preserving convexity of the objective function in the latent space.

\begin{figure*}
\includegraphics{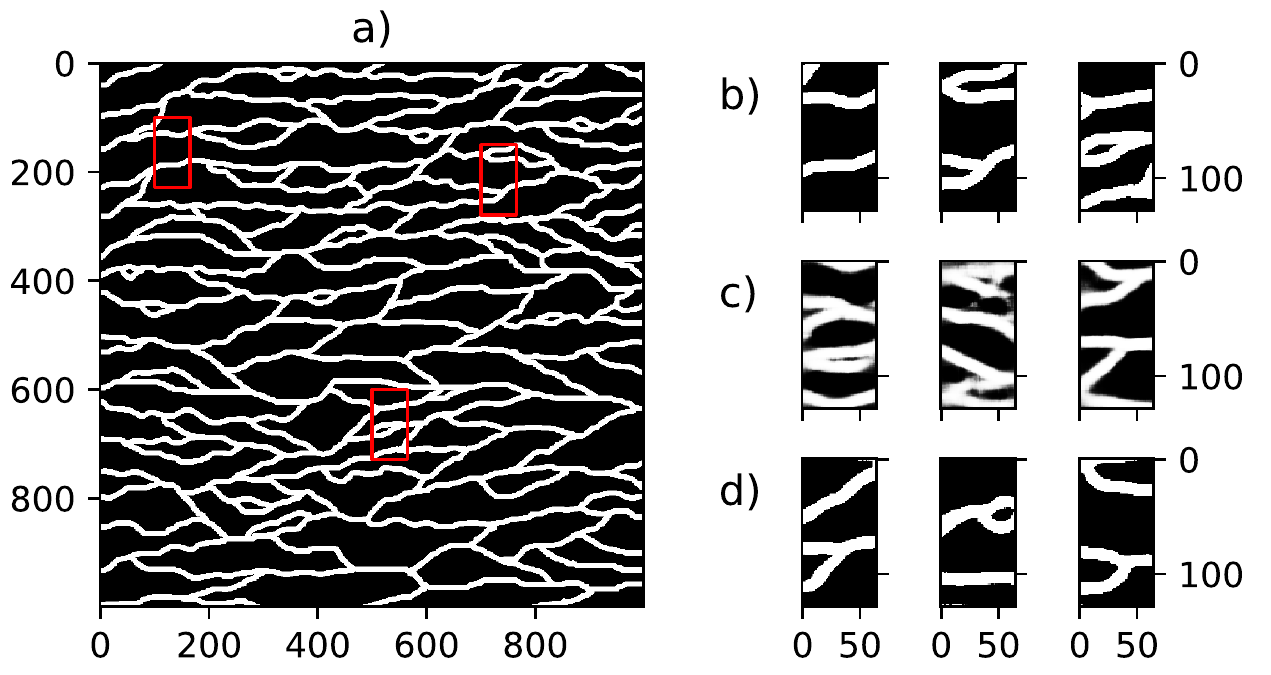}
\centering
\caption{(a) A 1000 $\times$ 1000 patch of the training image of Laloy et al. (2018), (b) cropped training samples whose location in (a) is shown red, (c) generated samples from our proposed VAE, and (d) generated samples from the SGAN proposed by Laloy et al. (2018).}
\label{Fig:ticropping}
\end{figure*}

To test our proposed method we implement a VAE in PyTorch \citep{paszke_automatic_2017} and use training samples cropped from a "training image" which is large enough to have many repetitions of the patterns at the cropping size---a requirement similar in MPS. For our synthetic case, we use the training image of 2500 $\times$ 2500 pixels from \citet{laloy_training-image_2018} and the cropping size is chosen to fit the setting of our synthetic experiment (explained in detail in Sec.\ \ref{Sec:Inverse}). Fig.\ \ref{Fig:ticropping}a shows a patch of the training image and the position of the three (cropped) training samples shown Fig.\ \ref{Fig:ticropping}b. Three generated samples from our proposed VAE trained with such croppings are shown Fig.\ \ref{Fig:ticropping}c. For comparison, Fig.\ \ref{Fig:ticropping}d shows three samples generated with the SGAN proposed by \citet{laloy_gradient-based_2019}. Patterns of generated samples in Fig.\ \ref{Fig:ticropping}c are not completely accurate comparing to those of the training image or the SGAN---they might display e.g.\ some breaking channels and smoothed edges. As mentioned above, this is expected for our proposed VAE because the approximate manifold fills some holes of the real manifold and may have less curvature. However, we argue that such inaccuracies may not cause significant error or bias while performing inversion in practice because an informative dataset will generally make the inversion land in appropriate models (given the prescribed patterns were selected correctly). More importantly, in contrast to the SGAN, a modified gradient-based inversion (such as that presented in Sec.\ \ref{Sec:SGDdec}) will generally find a consistent minimum when applied with our proposed VAE regardless of the initial model.

\subsection{Stochastic gradient descent with decreasing step size}
\label{Sec:SGDdec}

Note that even when topology is preserved and nonlinearity is contained, the data misfit function in the latent space might still present some local minima. Using our proposed VAE approach in the synthetic case study, the resulting misfit function seems to have the shape of a global basin of attraction with some local minima of less amplitude. To deal with such remaining local minima we propose to use a stochastic gradient descent (SGD) method instead of regular gradient-based optimization.

SGD methods are commonly used in training machine learning models to cope with large datasets \citep[e.g.][]{kingma_adam_2017} and it has also been shown they are able to find minima that are useful in terms of generalization \citep{smith_bayesian_2018}. They essentially use an estimator for the gradient of the objective function computed only with a batch of the data. Such estimator is used in each gradient descent iteration and may be written for the case of inversion in the latent space as:

\begin{equation}
\mathbf{z}_{k+1} = \mathbf{z}_{k} - \ell \cdot \nabla_{\mathbf{z}} \zeta(\mathbf{z})_k
\end{equation}

\noindent where $k$ denotes the iteration index, $\ell$ is the step size (or learning rate) and the gradient estimator $\nabla_{\mathbf{z}} \zeta(\mathbf{z})_k$ is computed by using Eq.\ (\ref{Eq:gradz}) for a data batch (i.e.\ a subset of $\mathbf{y}$) which is different for each $k$-th iteration but of constant size $b$. Relying on such estimator makes SGD methods less likely to get trapped in local minima if a sufficiently large step is chosen. However, if such a step is too large the optimization will have issues when it is close to the global minimum, usually seen in the form of high misfit and oscillations in the value of the objective function. Note that similar to other stochastic optimization methods, SGD only guarantees convergence to the global minimum with a certain probability, however if modified in the right way for the type of problems to be solved and its parameters chosen appropriately such probability could be very close to one.

Recently, it has been proposed that using SGD may be seen as optimizing a smoothed version of the objective function obtained by convolving it with the gradient "noise" resulting from batching \citep{kleinberg_alternative_2018}. The degree of noise (and therefore the degree of smoothness) is controlled by the ratio of the learning rate to the batch size $\ell / b$ \citep{chaudhari_stochastic_2018, smith_bayesian_2018}. Therefore if we choose to decrease the value of $\ell$ (while keeping $b$ constant) as the optimization progresses we might be able to achieve lower misfit values i.e.\ get sufficiently close the global minimum. This may be implemented by using:

\begin{equation}
\ell_{k+1} = c_{\ell} \cdot \ell_{k}
\label{Eq:decell}
\end{equation}

\noindent where a constant value of $c_{\ell} < 1.0$ and a starting value $\ell_{0}$ must be chosen. In practice, the method may be further improved by also decreasing the controlling (regularization) term in Eqs.\ (\ref{Eq:objz}) and (\ref{Eq:gradz}) in order to prevent that large initial steps diverge from the region of the latent space where the manifold is defined \citep{bora_compressed_2017}. Then, similarly to $\ell$ this may be done as:

\begin{equation}
\lambda_{k+1} = c_{\lambda} \cdot \lambda_{k}
\end{equation}

\noindent again a constant $c_{\lambda} < 1.0$ and a starting value $\lambda_{0}$ must be selected.

\begin{figure*}
\includegraphics[width=\textwidth]{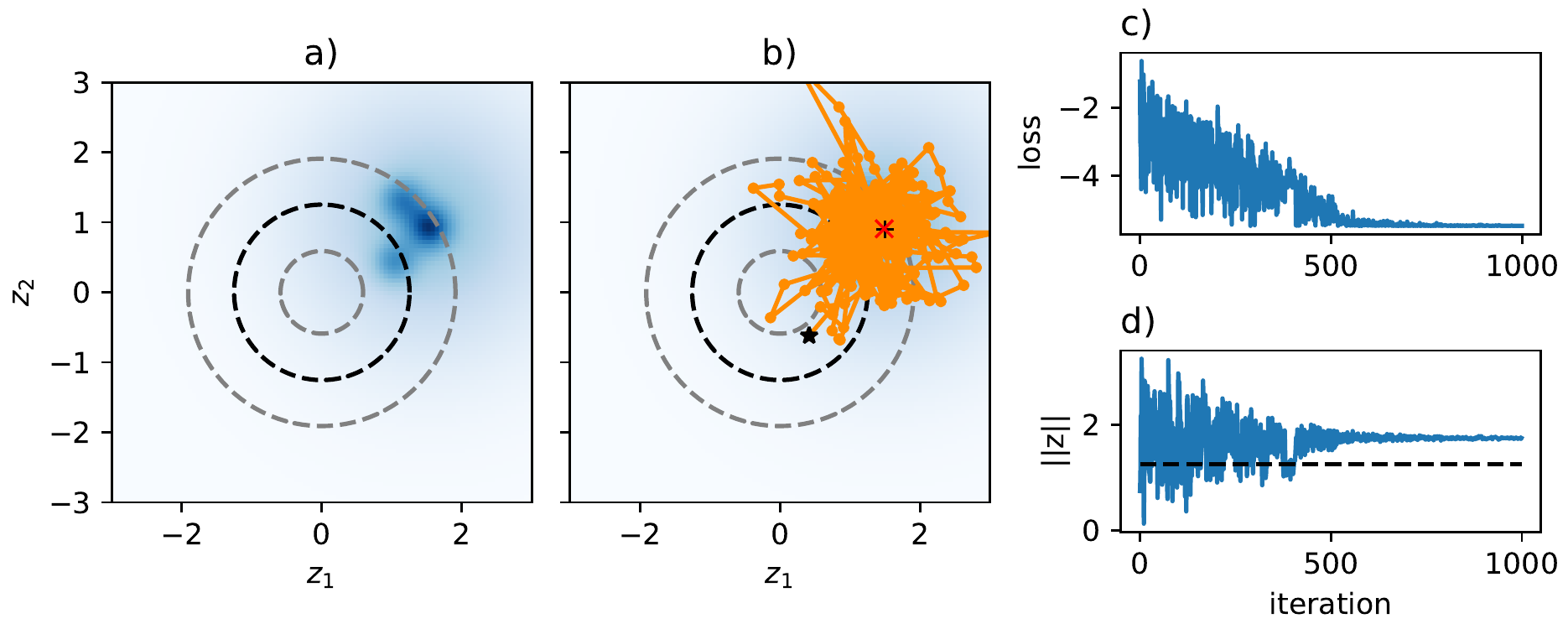}
\centering
\caption{Regularized gradient-based inversion in a synthetic two-dimensional latent space: (a) misfit (blue) and mean of $\chi$-distribution (black dashed) together with $16$- and $84$-th percentiles (gray dashed), (b) the same setting of (a) with an overlay of an instance of optimization (trajectory in orange) for a random initial model (black '$\star$'), showing also final model (red '$\times$') and true model (black '$+$'), (c) misfit vs.\ iteration number, and (d) norm of $\mathbf{z}$ vs.\ iteration number.}
\label{Fig:toysgd}
\end{figure*}

The combined effect of simultaneously decreasing $\ell$ and $\lambda$ is illustrated in Fig.\ (\ref{Fig:toysgd}) for a simple synthetic problem in a two-dimensional ($d=2$) latent space $\mathcal{R}^d$. The misfit term (i.e. first term of Eq.\ (\ref{Eq:objz})) of the synthetic problem is shown in Fig.\ \ref{Fig:toysgd}a. Assuming that $p(\mathbf{z})$ is a normal distribution $\mathcal{N}(\mathbf{0},I_d)$ where $I_d$ is a $d \times d$ identity matrix, we propose a specific regularization term $R(\mathbf{z})$ that will preferentially stay in the regions of higher mass (where most samples are located). This is done by radially constraining the search space by means of a $\chi$-distribution, i.e. the regularization term is written as:

\begin{equation}
R(\mathbf{z})= (\|\mathbf{z}\|-\mu_{\chi})^2
\label{Eq:ringreg}
\end{equation}

\noindent where $\mu_{\chi}$ is the mean for a $\chi$-distribution with $d$ degrees of freedom. Dashed lines in Fig.\ \ref{Fig:toysgd}a denote this mean together with the $16$- and $84$-th percentiles. In general, this is especially useful for higher dimensionalities where most of the mass of a normal distribution is far from its center \citep{domingos_few_2012}. Then, Eq.\ (\ref{Eq:objz}) may be rewritten as:

\begin{equation}
\zeta(\mathbf{z}) = \|\mathbf{f}(\mathbf{g}(\mathbf{z}))-\mathbf{y}\|^2 + \lambda (\|\mathbf{z}\|-\mu_{\chi})^2
\label{Eq:objzvae}
\end{equation}

\noindent and correspondingly Eq.\ (\ref{Eq:gradz}) may be expressed as:

\begin{equation}
\nabla_\mathbf{z}\zeta(\mathbf{z}) =  \mathbf{J}(\mathbf{z})^T \nabla_\mathbf{x}\gamma(\mathbf{x}) + 2\lambda\mathbf{z}\left(1-\frac{\mu_\chi}{\|\mathbf{z}\|}\right)
\label{Eq:gradzvae}
\end{equation}

\noindent As mentioned above, this gradient is often computed simply by adding a layer to the autodifferentiation of the generator. One optimization instance for a random initial model is shown in Fig.\ \ref{Fig:toysgd}b, while the behavior of the misfit and $\|\mathbf{z}\|$ is shown in Fig.\ \ref{Fig:toysgd}c,d. Notice the rather "noisy" inversion trajectory, but also its ability to escape local minima. The effect of decreasing $\ell$ is seen in Fig.\ \ref{Fig:toysgd}c by the decreasing of the oscillations amplitude as the optimization progresses, while the effect of decreasing $\lambda$ is noticeable in Fig.\ \ref{Fig:toysgd}d by the progressive shifting of $\|\mathbf{z}\|$ away from $\mu_\chi$.

The strategy described above and stated by Eq.\ (\ref{Eq:objzvae}) is generally applicable to DGMs that use an independent normal distribution as its probability distribution $p(\mathbf{z})$ and whose generator is well-behaved. In this work, we consider a VAE whose training parameters $\beta$ and $p(\bm{\epsilon})$ are chosen so that it results in a mildly nonlinear inversion for which such SGD strategy is generally useful.

\subsection{Inverse problem: traveltime tomography}
\label{Sec:Inverse}

To test our proposed method and compare it with a previous instance of inversion with a DGM, we consider an identical setting to that used in \citet{laloy_gradient-based_2019}. Such setting considers a dataset of borehole ground penetrating radar (GPR) traveltime tomography. To obtain a subsurface model $\mathbf{x} \in \mathbb{R}^D$ this method relies in contrasts of electromagnetic wave velocity which is related to moisture content and therefore to porosity for saturated media. The tomographic array considers a transmitter antenna in one borehole and a receiver antenna in the other, each of which is moved to different positions and a vector of measurements $\mathbf{y} \in \mathbb{R}^Q$ is obtained by taking the traveltime of the wave's first arrival for each transmitter-receiver combination. We assume that the sensed physical domain is a 6.5 $\times$ 12.9 m plane (i.e.\ the two-dimensional region between the boreholes) and is discretized in 0.1 $\times$ 0.1 m cells of constant velocity to represent spatial heterogeneity (i.e.\ a representation of $D=$ 65 $\times$ 129 $=$ 8385 cells is obtained). We consider a binary subsurface (e.g.\ composed of two materials with different porosity) with respective wave velocities of 0.06 and 0.08 m ns\textsuperscript{-1}. Measurements are taken every 0.5 m in depth (the first being at 0.5 m and the last at 12.5 m) resulting in a dataset of $Q=$ 625 traveltimes. For one instance of our synthetic case, we add normal independent noise $\bm{\eta} \sim \mathcal{N}(\mathbf{0},\sigma^2 \cdot I_{Q})$ where $\sigma^2$ is the noise variance and $I_{Q}$ is a 625 $\times$ 625 diagonal matrix.

Similarly to \citet{laloy_gradient-based_2019}, we first consider a fully linear forward operator $\mathbf{f}$ for which raypaths are always straight, i.e.\ independent of the velocity spatial distribution. For this case Eq.\ (\ref{Eq:inversion}) may be rewritten as:

\begin{equation}
\mathbf{y} = \mathbf{F}\mathbf{x} + \bm{\eta}
\label{Eq:flinear}
\end{equation}

\noindent where $\mathbf{F}$ is a matrix of dimension $Q \times D$ in which a certain row contains the length of the raypath in each cell of the model for a certain transmitter-receiver combination. The corresponding gradient of the misfit $\nabla_{\mathbf{x}}\gamma(\mathbf{x})$ to be used in Eq.\ (\ref{Eq:gradzvae}) for the solution of the inversion is:

\begin{equation}
\nabla_{\mathbf{x}}\gamma(\mathbf{x}) = -2\mathbf{F}^T (\mathbf{y}-\mathbf{Fx})
\label{Eq:gradxlinear}
\end{equation}

We also consider the case of a more physically realistic nonlinear forward operator $\mathbf{f}$ (see Eq.\ (\ref{Eq:inversion})) for which raypaths are not straight. In particular, we consider a shortest path (graph) method which uses secondary nodes to improve the accuracy of the simulated traveltimes as proposed by \citet{giroux_task-parallel_2013} and implemented in PyGIMLi \citep{rucker_pygimli:_2017}. For this case, when inversion with Eq.\ (\ref{Eq:gradzvae}) is pursued, we linearize the forward operator $\mathbf{f}$ in order to compute the gradient:

\begin{equation}
\nabla_{\mathbf{x}}\gamma(\mathbf{x}) = -\mathbf{S}(\mathbf{x})^T(\mathbf{y}-\mathbf{f}(\mathbf{x}))
\label{Eq:gradxnonlinear}
\end{equation}

\noindent where is $\mathbf{S}(\mathbf{x})$ is the $Q \times D$ jacobian matrix of the forward operator whose elements are:

\begin{equation}
[\mathbf{S}(\mathbf{x})]_{i,j} = \frac{\partial f_i(\mathbf{x})}{\partial x_j}
\end{equation}

\noindent The elements of the jacobian $\mathbf{S}(\mathbf{x})$ are computed by the shortest path method and also represent lengths of raypaths. In contrast to the linear case, these have to be recomputed in every iteration. Both the nonlinear forward operator and the need for recomputing the jacobian result in higher computational cost compared to the linear operator.

The method proposed in Sec.\ \ref{Sec:SGDdec} to perform gradient-based inversion with a VAE should work for the linear forward operator because the nonlinearity in the inverse problem arises only due to the generator $\mathbf{g}(\mathbf{z})$ which is moderate when the latter is well-behaved. However, since the considered nonlinear forward operator in Eq.\ (\ref{Eq:gradxnonlinear}) is only mildly nonlinear (when contrast in velocities is not extreme), the same method may also provide good inversion results for this operator.

\section{Results}
\label{Sec:Results}

\subsection{Training of VAE}

As previously mentioned, our proposed method relies on a VAE whose training parameters are selected in order to improve gradient-based inversion. The training samples are the croppings detailed in Sec.\ \ref{Sec:VAEasDGM} whose dimensionality is $D=$ 8325 and we consider a latent code of dimensionality $d=$ 20. The probability distribution $p(\mathbf{z})$ is an independent multinormal distribution $\mathcal{N}(\mathbf{0},I_d)$ with $I_d$ an identity matrix of size 20 $\times$ 20. The architecture of the encoder and the decoder includes 4 convolutional layers, 2 fully-connected layers and instance normalization is used between each layer (details may be consulted in the associated code). The training parameters relevant to our proposed  method are chosen in the following way: (1) $\beta$ in Eq.\ (\ref{Eq:aproxELBO}) is given a value of 1000, chosen by visually assessing the generated samples which also coincided with a moderate visual deviation from the prescribed $p(\mathbf{z})$; (2) the distribution $p(\bm{\epsilon})$ in Eq.\ (\ref{Eq:pepsilon}) is given the typical value of a diagonal unit variance ($\alpha=1.0$), which enforces local orthogonality while passing through the generator and therefore aids in preserving topology and controlling non-linearity. The value of $\beta$ is rather high compared to previous studies, e.g.\ \citet{laloy_inversion_2017} used a value of 20 for similar two-dimensional patterns, but seems to cause slightly higher compression ratios (in our work the compression ratio is 420 compared to 200 in the mentioned study). The VAE is trained by maximizing the lower bound in Eq.\ (\ref{Eq:aproxELBO}) using 10\textsuperscript{5} iterations and batches of 100 random croppings in each iteration (a GPU was used in order to reduce training time). In the following, we test the performance of this VAE when used for our proposed SGD-based inversion with a linear and a mildly nonlinear forward operator.

\subsection{Case with a linear forward model}

In this section, we consider the linear operator in Eq.\ (\ref{Eq:flinear}) and assess the performance of our proposed DGM inversion approach: using a VAE trained as above (to have a well-behaved generator) and SGD with both decreasing step size and regularization to optimize Eq.\ (\ref{Eq:objzvae}). We aim to show that such approach is robust regarding its convergence to the global minimum and therefore assess its performance by using 100 different initial models. Compared to previous studies, our proposed approach involves changes in both the DGM and the optimization, therefore we compare with the base cases listed in Table \ref{Tab:InvCases} to show the impact of each change proposed. As denoted by the columns of this table, the different cases consider: (1) VAE and SGAN as DGMs, (2) SGD and Adam \citep{kingma_adam_2017} as stochastic optimizers, (3) data batching for computing the gradient $\nabla_{\mathbf{z}}\zeta(\mathbf{z})$, which basically means using SGD when batching and using (regular) gradient-descent when not batching, (4) regularization in the latent space, with "origin" being the one proposed in \citet{bora_compressed_2017} and "ring" the one proposed herein, and (5) decreasing of the step size (or learning rate). Our proposed approach is then labeled as "VSbrd". We also show the chosen values for the step size $\ell$ and its decreasing factor $c_\ell$ when applicable---for these cases the values of $\lambda$ = 10.0 and $c_\lambda$ = 0.999 are used. The number of iterations for inversion is set to 3000 for all cases. When data batching is used, the batch size $b$ is 25 (of a total of 625) and is sampled with no replacement, then the whole dataset is used every 25 iterations (i.e.\ the number of epochs is 120 for a total of 120 $\times$ 25 = 3000 iterations).  Note that we compare against the approach in \citet{laloy_gradient-based_2019}, where SGAN is used as DGM and Adam (gradient-descent with adaptive moments) are used to optimize the resulting objective function---this case is labeled "SAnnn" in Table \ref{Tab:InvCases}. We also consider the case where we apply our proposed SGD to the same SGAN (labeled as "SSbnd"). For both of these cases instead of regularization we use stochastic clipping in the latent space (Laloy et al. 2018, 2019) because a uniform $p(\mathbf{z})$ with finite support is used.

\begin{table*}[t]
\centering
\begin{tabular}{lccccccc}
Case & DGM & GD & Data batching & Regularization & Decreasing & $\ell$ & $c_\ell$ \\
\hline
VSnnn & VAE & SGD & no & none & no & 1e-4 & - \\
VSbnn & VAE & SGD & yes & none & no & 1e-4 & - \\
VSbod & VAE & SGD & yes & origin & yes & 1e-2 & 0.95 \\
VSbrd & VAE & SGD & yes & ring & yes & 1e-2 & 0.95 \\
SAnnn & SGAN & Adam & no & none & no & 1e-2 & - \\
SSbnd & SGAN & SGD & yes & none & yes & 1e-3 & 0.95 \\
\end{tabular}
\caption{Configuration of our proposed approach (VSbrd) and the base cases for comparison.}
\label{Tab:InvCases}
\end{table*}

\begin{figure*}
\includegraphics{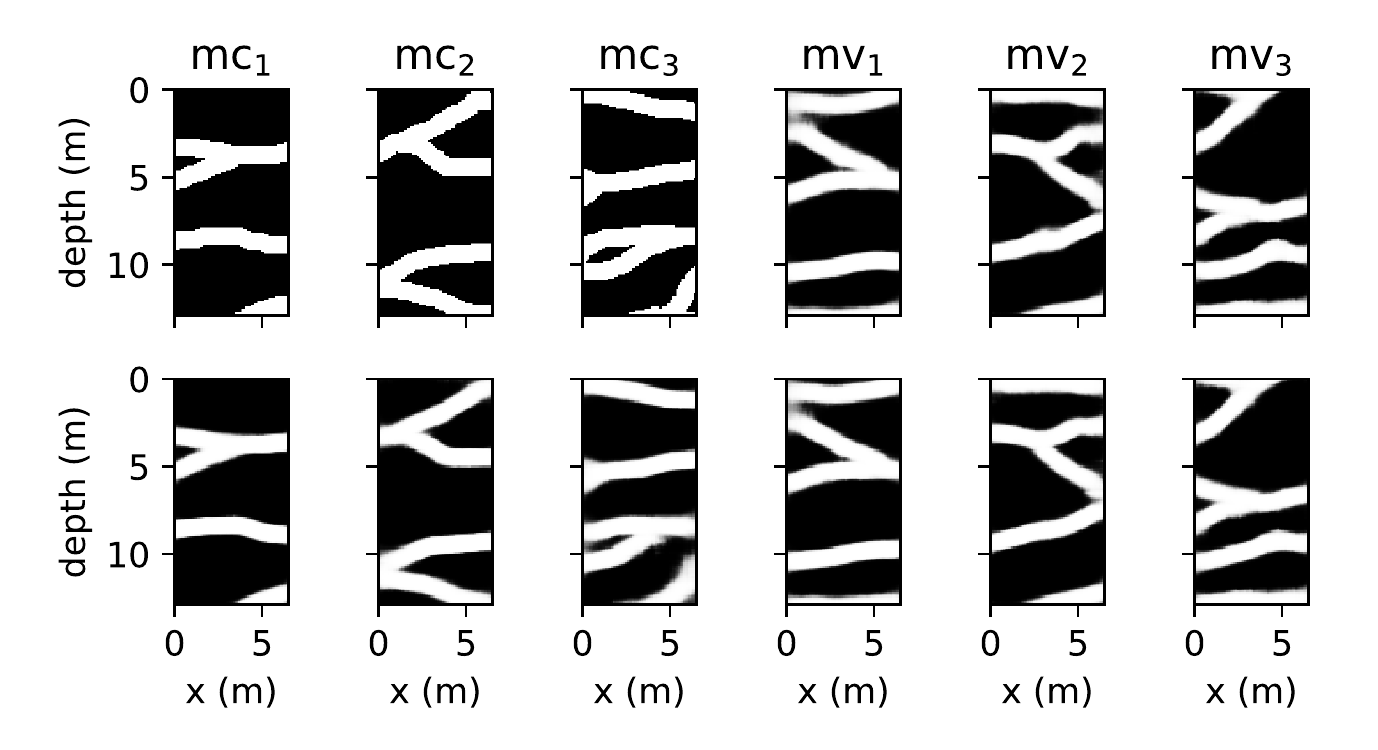}
\centering
\caption{Truth models (first row): cropped from training image (denoted by "mc") and generated from trained VAE (denoted by "mv"). Corresponding models resulting from encode-decode of truth models (second row). Subindex indicates level of complexity, with "1" being the least complex.}
\label{Fig:encdec}
\end{figure*}

We consider 6 different truth subsurface models to assess our method and compare with the base cases: (1) a set of three models cropped directly from the training image and (2) a set of three models obtained by generating from the trained VAE. Both sets include models with three different degrees of complexity. These truth models are shown in the first row of Fig.\ \ref{Fig:encdec} where "mc" refers to the first set, "mv" refers to the second set and the degree of complexity is denoted by a subscript, where "1" denotes least complex and "3" most complex. For the first set (mc), to avoid "memorizing" the croppings we exclude them (and any overlapping cropping) from the samples used to trained the VAE. The second set (mv) is similar to the one used by \citet{laloy_gradient-based_2019} to test the performance of their setup, only in their case the models were generated from a SGAN instead of a VAE. For each one of these truths, we generate synthetic data applying the forward operator $\mathbf{F}$ and use these data to perform gradient-based inversion for each case in Table \ref{Tab:InvCases}. 

We first consider no added noise to the synthetic dataset, hence after inversion the data misfit should be close to zero for inverted models that are sufficiently close to the global minimum. To define a threshold for this data misfit beyond which inverted models are "accepted", we use the RMSE between these synthetic data and data obtained by applying the forward operator on models resulting from passing the truth models through a VAE's encoding-decoding (these models are shown in the second row of Fig.\ \ref{Fig:encdec} and the corresponding values for the threshold are shown in Table \ref{Tab:thresholdRMSE}). This is done because we found the encode-decode reconstructed models to be visually very similar to the truth models (compare first and second rows of Fig.\ \ref{Fig:encdec}) and also show a low model RMSE when compared to them (computed just as the difference of pixel values between truth model and the encode-decode model and shown Table \ref{Tab:thresholdRMSE}). Once such threshold is defined for each truth model, gradient-based inversion is run for the same 100 initial models for all cases in Table \ref{Tab:InvCases}. Note that no convergence criteria were set in order to compare to all base cases (some cases such as "SAnnn" do not allow for easily defining such criteria) but in practice it is possible to set them for our proposed approach (VSbrd) in terms of a minimal change in either step size and/or data misfit. This also means that for some cases (including our proposed VSbrd) the 3000 iterations may not be necessary for all truths and all initial models. Results for the number of accepted inverted models are shown in Table \ref{Tab:Nacc} while the corresponding mean of the misfit (expressed as RMSE) for the 100 inversions is shown in Table \ref{Tab:meanDRMSE}.

\begin{table}
\centering
\begin{tabular}{lcc}
& data RMSE (ns) & model RMSE (-) \\
\hline
mc$_1$ & 0.606 & 0.104 \\
mc$_2$ & 0.998 & 0.147 \\
mc$_3$ & 1.096 & 0.173 \\
mv$_1$ & 0.524 & 0.057 \\
mv$_2$ & 0.958 & 0.098 \\
mv$_3$ & 0.734 & 0.085 \\
\end{tabular}
\caption{Data RMSE (ns) of encode-decode operation used to define thresholds (for the linear forward operator) and corresponding model RMSE.}
\label{Tab:thresholdRMSE}
\end{table}

\begin{table*}
\centering
\begin{tabular}{lcccccc|c}
& VSnnn & VSbnn & VSbod & VSbrd & SAnnn & SSbnd & VSbrd (noise) \\
\hline
mc$_1$ & 88 & 35 & 100 & 100 & 0 & 0 & 100\\
mc$_2$ & 96 & 50 & 100 & 100 & 0 & 0 & 100\\
mc$_3$ & 92 & 58 & 100 & 100 & 0 & 0 & 100\\
mv$_1$ & 100 & 75 & 100 & 100 & 2 & 0 & 100\\
mv$_2$ & 64 & 54 & 100 & 100 & 0 & 0 & 100\\
mv$_3$ & 91 & 71 & 100 & 100 & 0 & 0 & 100\\
\end{tabular}
\caption{Number of accepted inversions (using 100 different initial models) according to the defined threshold.}
\label{Tab:Nacc}
\end{table*}

\begin{table*}
\centering
\begin{tabular}{lcccccc|c|c}
& VSnnn & VSbnn & VSbod & VSbrd & SAnnn & SSbnd & threshold & VSbrd (noise) \\
\hline
mc$_1$ & 0.493 & 1.242 & 0.544 & 0.405 & 4.538 & 3.988 & 0.606 & 0.480\\
mc$_2$ & 0.620 & 1.525 & 0.690 & 0.546 & 5.266 & 4.495 & 0.998 & 0.593\\
mc$_3$ & 1.066 & 1.507 & 0.885 & 0.833 & 3.298 & 3.775 & 1.096 & 0.880\\
mv$_1$ & 0.456 & 0.963 & 0.121 & 0.062 & 4.777 & 4.703 & 0.524 & 0.273\\
mv$_2$ & 1.061 & 1.308 & 0.289 & 0.080 & 5.236 & 4.717 & 0.958 & 0.268\\
mv$_3$ & 1.282 & 1.233 & 0.224 & 0.033 & 4.770 & 4.774 & 0.734 & 0.256\\
\end{tabular}
\caption{Mean RMSE (ns) of inversions using 100 different initial models and defined threshold for accepting models.}
\label{Tab:meanDRMSE}
\end{table*}

As seen in Table \ref{Tab:Nacc}, given our defined threshold: (1) the cases where VAE and SGD with decreasing step were used (VSbod and VSbrd) resulted in all inverted models being accepted, (2) the cases where SGAN was used (SAnnn and SSbnd) resulted in almost all models being rejected (only two models accepted for SAnnn with truth mv\textsubscript{1}), and (3) the cases where VAE and non-decreasing step size SGD was used (VSnnn and VSbnn) resulted in some inverted models being accepted. Note also that using SGD (data batching) without a decreasing step size (VSbnn) results in less accepted models compared to GD (VSnnn), highlighting the importance of our proposed decreasing step size and regularization. As shown in Table \ref{Tab:meanDRMSE} a higher mean RMSE is related to a lower number of accepted models. Furthermore, two things worth highlighting in Table \ref{Tab:meanDRMSE} are (1) the general improvement for inversion with SGAN caused by our proposed SGD compared to Adam for most truth models (compare SSbnd and SAnnn), which means that our proposed SGD is advantageous regardless of the DGM, and (2) the slight improvement caused by our proposed regularization compared to the one from \citet{bora_compressed_2017}.

\begin{figure*}
\includegraphics[width=\textwidth]{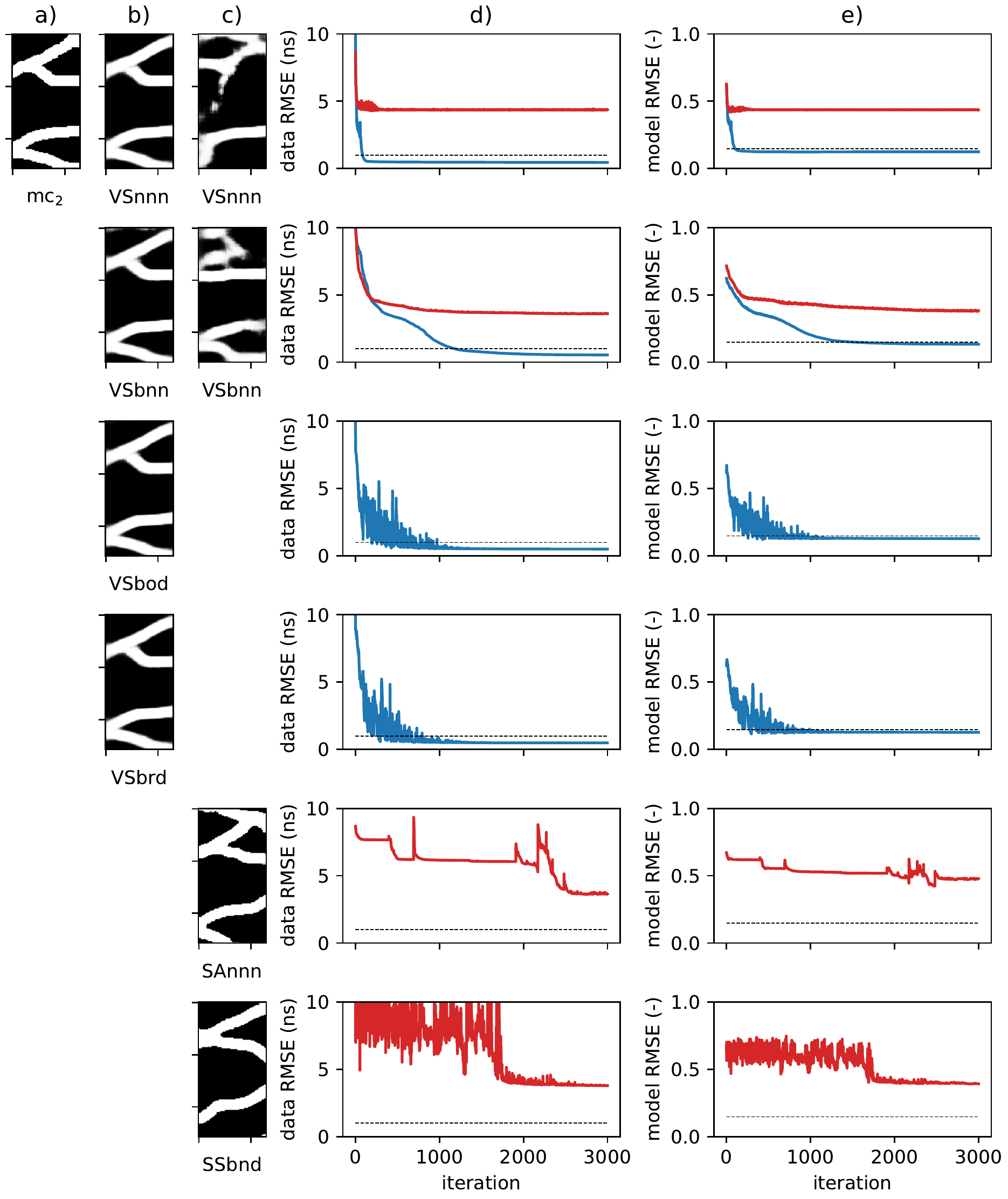}
\centering
\caption{Examples of inverted models for mc\textsubscript{2} truth for all cases in Table \ref{Tab:InvCases}: (a) accepted models according to defined threshold, (b) rejected models, (c) data RMSE vs.\ iterations plots (blue for accepted models and red for rejected models and dashed line indicates defined threshold) and (d) model RMSE vs.\ iterations plots (dashed line indicates model RMSE for encode-decode operation).}
\label{Fig:DGMsinv}
\end{figure*}

\begin{figure*}
\includegraphics[width=\textwidth]{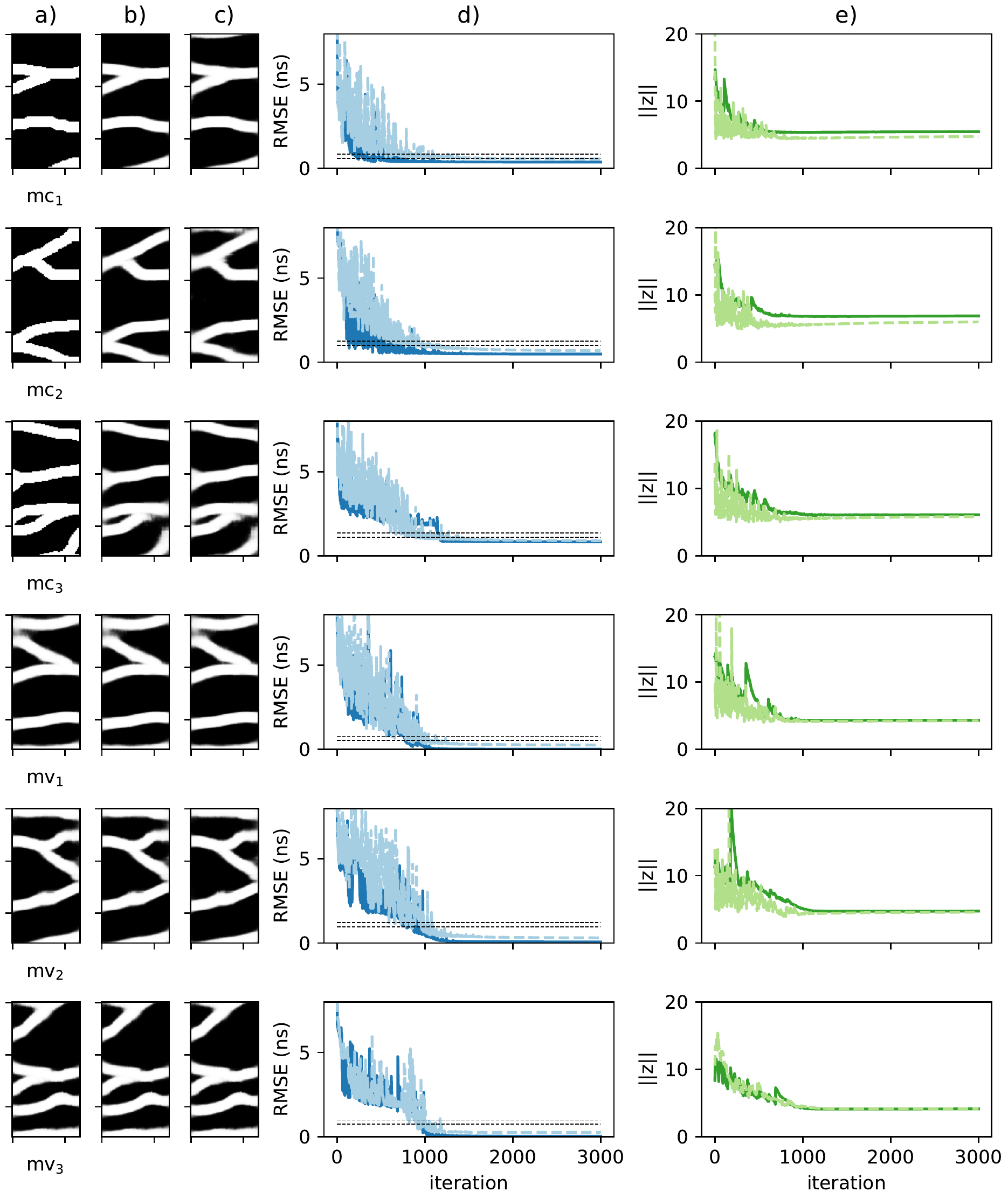}
\centering
\caption{Examples of gradient-based inversion using our proposed approach (VSbrd) for all truth models and the linear forward operator: (a) truth models, (b) inverted models with no added noise, (c) inverted models with added noise, (d) RMSE vs.\ iterations plots (no noise case in dark blue and noise case in light blue; lower dashed line indicates the defined threshold while upper dashed line is threshold plus $\sigma=0.25$), and (c) norm of $\mathbf{z}$ vs.\ iterations plots (no noise case in dark green and noise case in light green).}
\label{Fig:VSbrdinv}
\end{figure*}

Examples of inverted models obtained for the different cases in Table \ref{Tab:InvCases} using the cropped truth with moderate complexity (mc\textsubscript{2}) are shown in Fig.\ \ref{Fig:DGMsinv}. Here, truth models are shown in Fig.\ \ref{Fig:DGMsinv}a while Fig.\ \ref{Fig:DGMsinv}b shows one example of an accepted model for cases that have at least one (VSnnn, VSbnn, VSbod and VSbrd). Similarly, Fig.\ \ref{Fig:DGMsinv}c shows one example of a rejected model for applicable cases (VSnnn, VSbnn, SAnnn and SSbnd). Finally, the corresponding data RMSE vs.\ iteration number plots are shown in Fig.\ \ref{Fig:DGMsinv}d (in blue for accepted models and red for rejected ones) and corresponding model RMSE plots are shown in Fig.\ \ref{Fig:DGMsinv}e. Note both the higher similarity with the truth model (i.e.\ note the low model RMSE and compare models in Fig.\ \ref{Fig:DGMsinv}b,c with those in Fig.\ \ref{Fig:DGMsinv}a) and the lower RMSE for accepted models. Also, examples of inverted models for our proposed approach (VSbrd) using all the truths are shown in Fig.\ \ref{Fig:VSbrdinv}b, together with plots of RMSE vs.\ iteration number (Fig.\ \ref{Fig:VSbrdinv}d) and norm of $\mathbf{z}$ vs.\ iteration number (Fig.\ \ref{Fig:VSbrdinv}e). For cropped truths (mc) it seems that visual similarity decreases and final data RMSE of inverted models increases as complexity increases, whereas for generated truths they seem independent of complexity. Notice the overshoot in $\|\mathbf{z}\|$ in the initial iterations and its eventual convergence close to $\mu_\chi$ as defined in Eq.\ (\ref{Eq:ringreg}).

To study the effect of noise for our proposed approach (VSbrd), we added noise with a standard deviation $\sigma$ = 0.25 ns to the synthetic traveltime data. Corresponding results are shown in the rightmost column of Tables \ref{Tab:Nacc} and \ref{Tab:meanDRMSE} and in Fig.\ \ref{Fig:VSbrdinv}c (with corresponding data RMSE and $\mathbf{z}$ norm plots in Fig.\ \ref{Fig:VSbrdinv}d,e). The threshold in this case is set equal to the one for the noise-free case plus $\sigma$ and when using it all inverted models with our proposed approach are accepted. It is also worth noticing the relative robustness of the method to noise, as shown by the corresponding mean misfit values in Table \ref{Tab:meanDRMSE} that indicate no significant overfitting, i.e. the mean misfit values are close to the noise-free threshold plus $\sigma$ even if no traditional regularization was used. The latter means that optimizing in the latent space of the DGM is effectively constraining the inverted models to display the prescribed patterns.

\subsection{Case with a nonlinear forward model}

\begin{figure*}
\includegraphics[width=\textwidth]{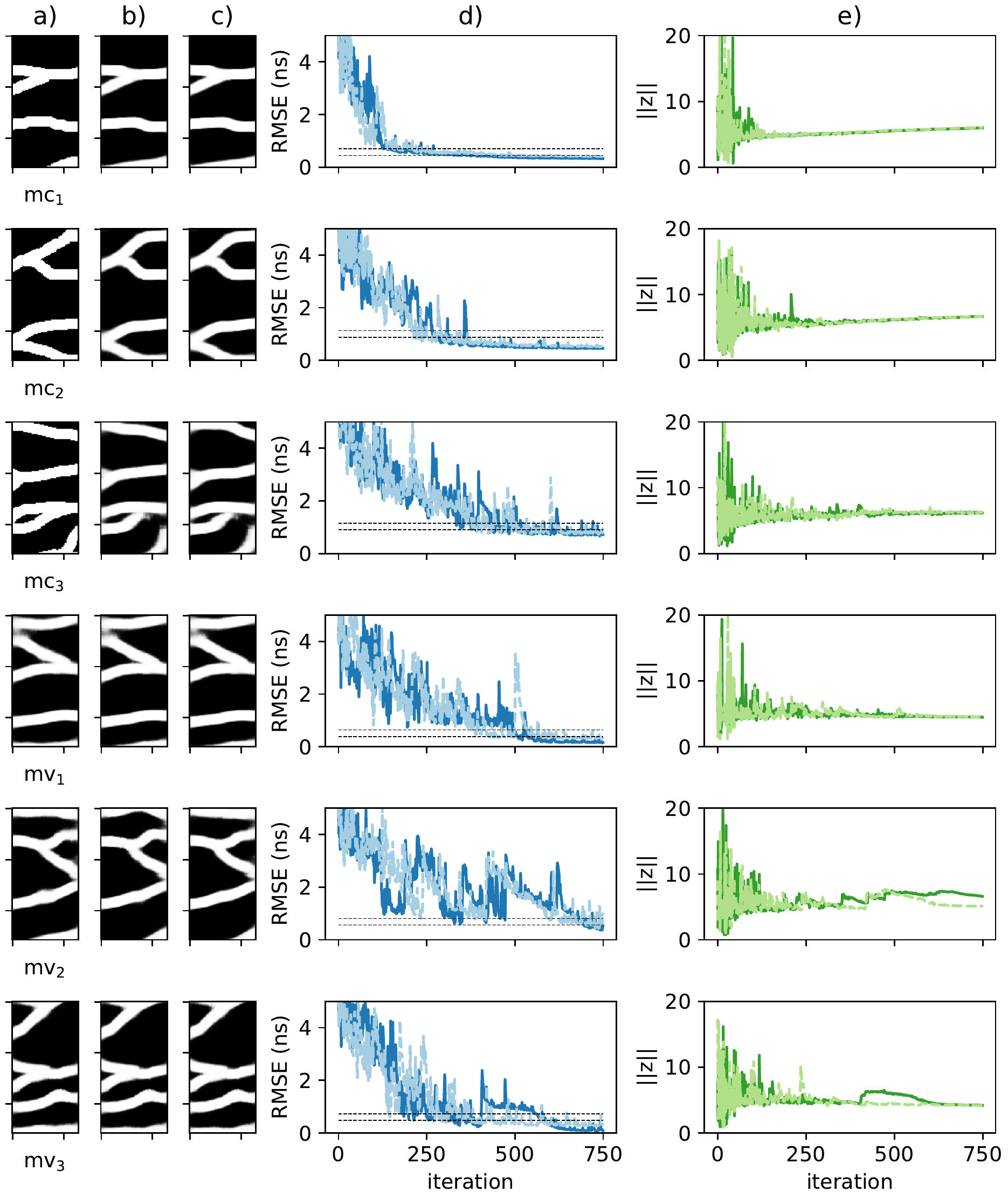}
\centering
\caption{Examples of gradient-based inversion using our proposed approach (VSbrd) for all truth models and the nonlinear forward operator: (a) truth models, (b) inverted models with no added noise, (c) inverted models with added noise, (d) RMSE vs.\ iterations plots (no noise case in dark blue and noise case in light blue; lower dashed line indicates the defined threshold while upper dashed line is threshold plus $\sigma=0.25$), and (c) norm of $\mathbf{z}$ vs.\ iterations plots (no noise case in dark green and noise case in light green).}
\label{Fig:VSbrdinvnonlin}
\end{figure*}

After showing that our proposed method works with the linear forward operator for the synthetic case considered, we now test its performance with a nonlinear forward operator. For inversion, the general form of Eq.\ (\ref{Eq:inversion}) is used and the gradient in the latent space given in Eq. (\ref{Eq:gradzvae}) is computed using Eq.\ (\ref{Eq:gradxnonlinear}). As mentioned in Sec.\ \ref{Sec:Inverse}, we consider a shortest path method to solve for the traveltime for which we use 3 secondary nodes added to the edges of the velocity grid. Note that the jacobian $\mathbf{S}(\mathbf{x})$ in Eq.\ (\ref{Eq:gradxnonlinear}) has to be recomputed at every iteration. Given the higher computational demand for inversion with the nonlinear forward operator and since it was already shown to be the best performing approach for the linear forward operator, we only test our proposed approach VSbrd with all the truths and for a single initial model (Fig.\ \ref{Fig:VSbrdinvnonlin}). This was done both without noise and with noise added using the same standard deviation $\sigma$ = 0.25 as in the linear operator scenario. We select the following values for the required inversion parameters: $\ell=$ 0.1, $c_\ell=$ 0.8, $\lambda=$ 1.0 and $c_\lambda=$ 0.99. The total number of iterations is 750 with data batching of size 25 similar to the linear case. Note that to further reduce the number of iterations required for inversion we use a lower $c_\ell$ compared to the linear case, but the decreasing in Eq.\ (\ref{Eq:decell}) is only done every 5 iterations. This may cause the method to converge to the global minimum with lower probability, however it seems to still be high enough since all of the inversions with no added noise are very similar to the truth models. Also, using the threshold obtained by encoding-decoding the truth models (now computed with the nonlinear forward operator) all inverted models are accepted (these models are shown in Fig.\ \ref{Fig:VSbrdinvnonlin}b). When considering added noise, results are similar but inversion seems to converge to the global minimum with slightly lower probability (6 out of 8 inversions are accepted) and accepted models are shown in in Fig.\ \ref{Fig:VSbrdinvnonlin}c. The behavior of the misfit during optimization (Fig.\ \ref{Fig:VSbrdinvnonlin}d) is similar to the linear case, although oscillations of a slightly higher amplitude are still visible in the last iterations (mainly due to the lower number of iterations). To partially solve the latter issue, we take as inverted model the model with lowest misfit and not the one for the final iteration (these are the models shown in Fig.\ \ref{Fig:VSbrdinvnonlin}b,c). The plot of the norm of $\mathbf{z}$ vs.\ iterations in Fig.\ \ref{Fig:VSbrdinvnonlin}e shows a similar behavior to the linear case, although there seems to be more oscillations in $\|\mathbf{z}\|$ during initial iterations.

\section{Discussion}
\label{Sec:Discussion}

When training the VAE for our considered synthetic case, only the selection of $\beta$ was done while the variance of $p(\bm{\epsilon})$ was not changed (a unity variance $\alpha=1.0$ was used). However, as noted in Sec. \ref{Sec:VAEasDGM} and in Fig.\ \ref{Fig:deformed}, the variance $\alpha$ also has an impact on (gradient-based) inversion as it might be used to control the generator's nonlinearity and its induced changes in topology.

In order to select SGD parameters in our proposed approach, we suggest looking jointly at the behavior of the misfit and norm of $\mathbf{z}$. For instance, if a certain number of iterations is desired for computational reasons, we suggest choosing first $\ell$ and $c_\ell$ that produce a behavior of the misfit similar to that in Fig.\ \ref{Fig:VSbrdinv}d, i.e.\ oscillations of high amplitude at the beginning and then progressive attenuation of the oscillations in such a way that at the end they are negligible. Note, however, that inversion may have to be run a few times because divergence may occur during initial iterations (this is easily seen in the value of $\|\mathbf{z}\|$ taking values far from $\mu_\chi$). Once $\ell$ and $c_\ell$ are chosen, the selection of $\lambda$ and $c_\lambda$ is done only to prevent divergence, this may be achieved by looking for a behavior similar to that in Fig.\ \ref{Fig:VSbrdinv}e. An initial overshoot in $\|\mathbf{z}\|$ is normal (and even necessary) since the method is exploring more rapidly the latent space, however, it should eventually converge to a value close to $\mu_\chi$.

The results for gradient-based inversion using our proposed approach point to a (possible) conflict between the accuracy of the reproduced patterns and the feasibility of gradient-based inversion with DGMs. As mentioned above, this is due to a non-convex objective function in latent space resulting from the generator's nonlinearity and its induced changes in topology. In this work, we argue that nonlinearity and changes in topology might be safely controlled by selecting certain values of $\alpha$ and $\beta$ while training a VAE in order to improve performance of gradient-based inversion. We empirically prove the validity of this statement for our case study with specific values $\alpha$ and $\beta$. In general (for inversion with DGMs), this implies that a tradeoff between generative accuracy and a well-behaved generator may be found. The latter statement also supports our assumption regarding the "holes" of the real manifold for the case of channel patterns (as mentioned in Section \ref{Sec:VAEasDGM}): when approximating the real manifold using a VAE with a well-behaved generator, the approximate manifold will tend to fill the holes and therefore produce breaking channels. While the generator's nonlinearity was already identified by \citet{laloy_gradient-based_2019} as a potential factor for hindering gradient-based inversion, its causes (curvature and topology of the real manifold) and the possible induced changes in topology have not been previously explained as factors in degrading the performance of gradient-based inversion in the latent space (to the authors' knowledge).

In general, good performance of DNNs for some tasks is usually associated with their ability to change topology \citep{naitzat_topology_2020}. However, when one wants to use the latent variables or codes of DGMs for further tasks and not just for generation, these changes in topology might become an issue. For instance, we interpret the misfit "jumps" seen in gradient-based inversion with SGAN (as seen in Fig.\ \ref{Fig:DGMsinv}c for case SAnnn) as resulting from the "gluing" or "collapsing" in latent space of holes in the real manifold---either caused by an induced change in topology or a high nonlinearity in the SGAN generator. Some studies have even suggested that if one wants to obtain useful geometric interpretations in the latent space (e.g.\ to perform interpolation), the activation functions should be restricted to ones that are smooth \citep{shao_riemannian_2017, arvanitidis_latent_2018}, that means e.g.\ not using the ReLU activation function that is generally recognized to result in faster learning. In contrast, in this work we do consider ReLU activation functions but control the changes in topology by means of a combination of $\alpha$ and $\beta$, whether this might nullify the advantages of ReLU is still an open question. Note  however that, in general, control of induced changes in topology and high nonlinearities (as in our proposed approach) might be useful for any inversion method that relies in the concept of a neighborhood (e.g.\ MCMC and ensemble smoothers).

Besides its good performance for gradient-based inversion, a further advantage of our approach when compared to the previous approaches is that when the data used for inversion is not sufficiently informative, regularization in the latent space might be used to constrain to the most common patterns with our regularization term in Eq.\ (\ref{Eq:ringreg}). This statement provides an interesting paradigm where regularization in latent space might be seen as a flexible way to incorporate complex regularization. In contrast, a disadvantage of our proposed approach is that GANs in general result in higher generative accuracy (all generated patterns look more similar to those in the training image), however, as previously mentioned this may be in conflict with performance of certain inversion methods. Also, as may be noticed in the relation between the data misfit and the degree of complexity for cropped truths, a limiting factor in using our VAE is its inability to produce new highly complex patterns. However, this lack of innovation (or sample diversity) is generally present in other methods and may be even more severe for regular GANs, where the phenomenon is known as mode collapse. Recently, different ways to control such mode collapse in VAEs and GANs have been proposed \citep{metz_unrolled_2017, salimans_improved_2016}.

Finally, regarding our proposed SGD we must note that similar results might be obtained with a MCMC method where information about the gradient is taken into account. For example, \citet{mosser_stochastic_2018} use a Metropolis-adjusted Langevin method which basically follows a gradient-descent and adds some noise to the step. However, the noise added to the gradient step in our approach is different---SGD noise has been shown to be approximately constant but anisotropic \citep{chaudhari_stochastic_2018}. Another possible alternative to our method is to use Riemannian optimization, which is possible when the DGM approximate manifold is smooth. Although it is possible to compute the direction of the gradient by using the pullback Riemannian metric, which may be obtained as suggested by e.g. \citet{shao_riemannian_2017, chen_metrics_2018, arvanitidis_latent_2018}, it is not straightforward to compute the step because it would have to be along a geodesic curve instead of a straight path and such geodesics are computationally demanding to compute.

\section{Conclusions}
\label{Sec:Conclusions}

In this work the principal difficulties of performing inversion with deep generative models (DGMs) are reviewed and a conflict between generated pattern accuracy and feasibility of gradient-based inversion is identified. Also, an approach based on a variational autoencoder (VAE) as DGM and a modified stochastic gradient descent method for optimization is proposed to address such conflict. We show that two training parameters of the VAE (the weight factor $\beta$ and the variance $\alpha$ of the encoder's noise distribution $p(\bm{\epsilon})$) may be chosen in order to obtain a well-behaved generator $\mathbf{g}(\mathbf{z})$, i.e.\ one that is mildly nonlinear and approximately preserves topology when mapping from latent space to ambient space. This helps in maintaining the convexity of the misfit function in the latent space and therefore improves the behavior of gradient-based inversion. We highlight changes in topology which have not been previously identified as impacting the convexity of the inversion objective function. In contrast to prior studies where gradient-based inversion was used, our approach converges to the neighborhood of the global minimum with very high probability for both a linear forward operator and a mildly nonlinear forward operator with and without noise. We argue that when using DGMs in inversion, a tradeoff may be found where inverted models are close enough to the prescribed patterns while low cost gradient-based inversion is still applicable---our proposed approach relies on such tradeoff and produces inverted models with significant similarity to the training patterns and a low data misfit.

\section*{Acknowledgements}

This work has received funding from the European Union’s Horizon 2020 research and innovation program under the Marie Sklodowska-Curie grant agreement number 722028 (ENIGMA ITN).

\section*{Computer code availability}

All code necessary to reproduce the test case is available at: \href{https://github.com/jlalvis/VAE_SGD}{https://github.com/jlalvis/VAE\_SGD}.



\bibliography{Paper_VAESGD}

\end{document}